2017

# New Method for Keyword Extraction from Patent Claims

EXTRACTING MEANINGFUL WORDS
ROSSI JULIEN

EUROPEAN PATENT OFFICE | Rijswijk / Den Haag
AMSTERDAM BUSINESS SCHOOL | Amsterdam
UNIVERSITEIT VAN AMSTERDAM | Amsterdam

# Master Thesis

| | |
|---|---|
| **Student** | Julien ROSSI<br>Amsterdam Business School<br>Student ID 11151056<br>MBA Big Data & Business Analytics |
| **Promotor** | Evangelos KANOULAS<br>UvA |
| **Co-Promotor** | Marc SALOMON<br>UvA, Amsterdam Business School |
| **EPO Supervisor** | Matthias WIRTH<br>European Patent Office |

# Acknowledgements

The author wishes to thank Matthias Wirth for supervising, monitoring, advising and making this work happen.

Thanks to Evangelos Kanoulas for advising, instructing and challenging me every week.

Thanks to Domenico Golzio and Stefan Klocke for allowing me in their teams. Thanks to Delphine Pouilley and Andreas Stabel for connecting me with the right people when I was searching a thesis.

Thanks to Volker Hähnke for setting up and delivering the test results.

# 1 INTRODUCTION

The legal field of intellectual property is the cornerstone of the "Knowledge Economy" (Drucker, 1969), our current economic era where "Production and services [are] based on knowledge-intensive activities […] a greater reliance on intellectual capabilities […]"[1], as it allows companies to claim and manage rights to the economic value of the innovations they produce.

Patents are one form of intellectual property rights that legally protect the monopoly of an invention. The recent high-profile raging "Patent Wars"[2] made clearly visible how strategically central the ownership of patents was for innovative companies, and how dedicated they are to maintaining their monopoly.

In this context, litigation arises from two different approaches:

- **Infringement on existing patents:** when a party has evidence that another party used a patented invention without permission. This is the enforcement of the monopoly rights, as granted by the patent.
- **Invalidation of existing patent or patent application:** when a party has evidence that an invention described in a patent or in an application for a patent, does not have sufficient merits to be granted a protection. An invalidated patent translates into a decrease of future cash-flows for the owner.

Invalidation is a challenge for the owners of patent portfolios: it is a core competence of patent offices around the world, as the quality and thoroughness of their work influences the likelihood of future invalidations. As governmental organizations, they have a duty to the public to properly identify and defend innovation, and not grant monopoly rights based on improper or weak claims of innovation.

We will focus in this paper on the task of search for prior art in patent literature, we will review the importance of this task in the financial and organizational environment of the European Patent Office (EPO), assess the existing academic research and literature on that topic, explain the method we demonstrate, and evaluate how it compares to the standards in place and discuss its merits.

We will also show how improving search efficiency integrates in a wider business strategy of the European Patent Office.

---

[1] Walter W. Powell, Kaisa Snellman, "The Knowledge Economy", 2004 in "Annual Review of Sociology"
[2] See the economic press at large about the "Smartphone Patent Wars", involving Apple, Samsung, Google, and many others. See https://en.wikipedia.org/wiki/Smartphone_patent_wars for a quick overview. The "Internet of Things" is currently considered as the next battleground, see https://prescouter.com/2016/09/iot-patent-war/ for an introduction

# 2 THE WORLD OF PATENTS

## 2.1 WHAT IS A PATENT?

A patent is the legal right to a monopoly of the usage or the sale of an invention. It is granted to a party (a corporation, most of the time), for a limited extent of time, and applies to a limited geographical area. It is granted by the patent office who has authority on the territory where the inventor seeks protection.

## 2.2 THE ROLE OF PATENT OFFICES

National and regional offices have full authority over their territory to apply the local patent law. The European Patent Office administers Europe-wide[3] patents, which represent the same right as a national patent delivered in each of the participating countries.

A patent office receives application documents, produces a search report, where it evaluates whether the application is actually novel, inventive, and industrially applicable. If so, it proceeds with the formal step of granting a patent to the applicant. The search establishes what the existing prior art in the field of the invention is, and evaluates if the invention is actually novel and acceptable.

PCT filings are applications that can be filed in any of 22 identified Patent Offices[4]. After an initial search, these filings can be transmitted to other Patent Offices for the purpose of receiving a patent corresponding to the filing, valid in their respective authority area. Applicants are free to select any of the 22 offices for the initial application.

## 2.3 CONTENTS OF A PATENT DOCUMENT

The different parts of a patent document comply with the rules of the European Patent Convention (EPC):

- Description (Rule 42 EPC): a specific scientific technical description of the invention, with drawings, figures, tables
- Claims (Rule 43 EPC): a legal description of what is legally protected in the invention
- Classification: one of the most used classification system is the Cooperative Patent Classification (CPC). Each patent can be allocated multiple numbers of CPC[5] classes. The CPC tree starts with 8 domains identified by letters A-H, it is then split into more than 600 subclasses, each refined in more details. This results in more than 250000 final classification symbols. The classification is a task performed by the patent offices.

---

[3] Countries part of the European Patent Convention
[4] See https://en.wikipedia.org/wiki/Patent_Cooperation_Treaty
[5] http://www.cooperativepatentclassification.org/index.html

# 3 ORGANIZATIONAL ENVIRONMENT AND CHALLENGES

## 3.1 THE EUROPEAN PATENT OFFICE

The European Patent Office (hereafter referred as EPO) is the executive body for the European Patent Organization. It grants European patents for the contracting states to the European Patent Convention (EPC), a multilateral treaty signed in October 1973, and later amended to extend the list of contracting states. It was recently revised in 2000.

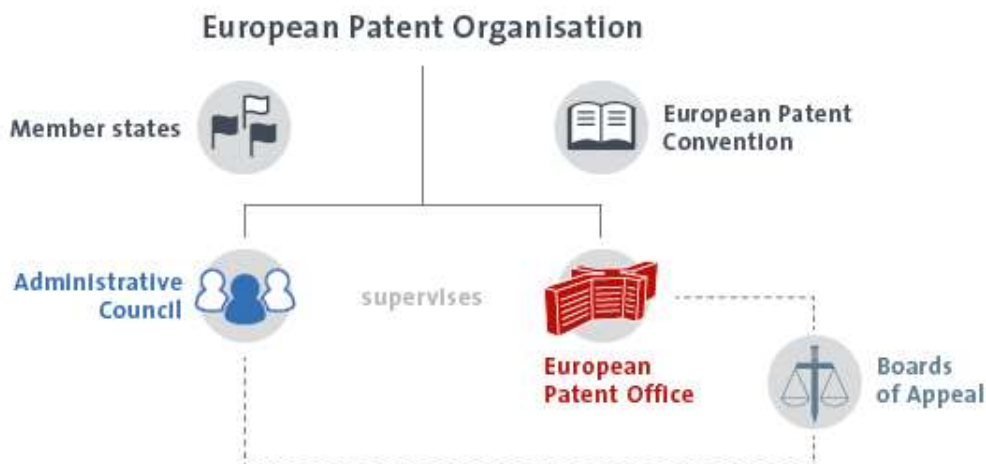

The EPO is a non-profit organization, whose revenues come from the collection of fees in exchange of processing application and continuing legal protection of granted patents. It is headquartered in Munich (Germany), with offices in Den Haag/Rijswijk (Netherlands), Berlin (Germany), Vienna (Austria) and Brussels (Belgium). It employs approximately 6800 people, including 4300 patent examiners[6].

It is the duty of the EPO to implement a patent grant process compliant with the articles of the EPC, that process being the backbone of the organization

## 3.2 FINANCES AND REVENUE MODEL

The yearly financial statements[7] are established and certified by an auditor, they follow the IFRS standards.

The EPO collects processing fees from applicants at various stages of the patent lifecycle, corresponding to the performance of specific tasks and decisions, or the continuous performance of protection of one's IP property rights. The EPO has only limited influence on the amount of fees, and cannot expect to be able to raise them significantly or to implement another revenue model. Projections are made in order to forecast future revenues, based on economic and industrial cycles.

The EPO has no social capital, so Equity is limited to retained earnings, an account with a negative balance, due to the compounding of accounting losses over years. The EPO does not own any debt.

---

[6] EPO, "Social Report 2016" www.epo.org/service-support/publications.html (visited July 2017)
[7] See https://www.epo.org/service-support/publications.html?pubid=19#tab3 (visited July 2017)

The striking element is the importance of actuarial adjustments performed on long-term liabilities (defined benefits obligations, e.g. pensions and benefits, called RFPSS). Since 2010, IFRS is stating that organizations should report variations of the NPV of those liabilities, as recalculated with currently observed interest and discount rates, while their terminal values are highly sensitive to the volatility of the risk-free rate. These adjustments caused accounting profits and losses amounting to 3 to 8 billion €, while the 'ordinary' profits and losses are in the 2 to 3 hundreds of millions of €. These adjustments do not result in cash-flows.

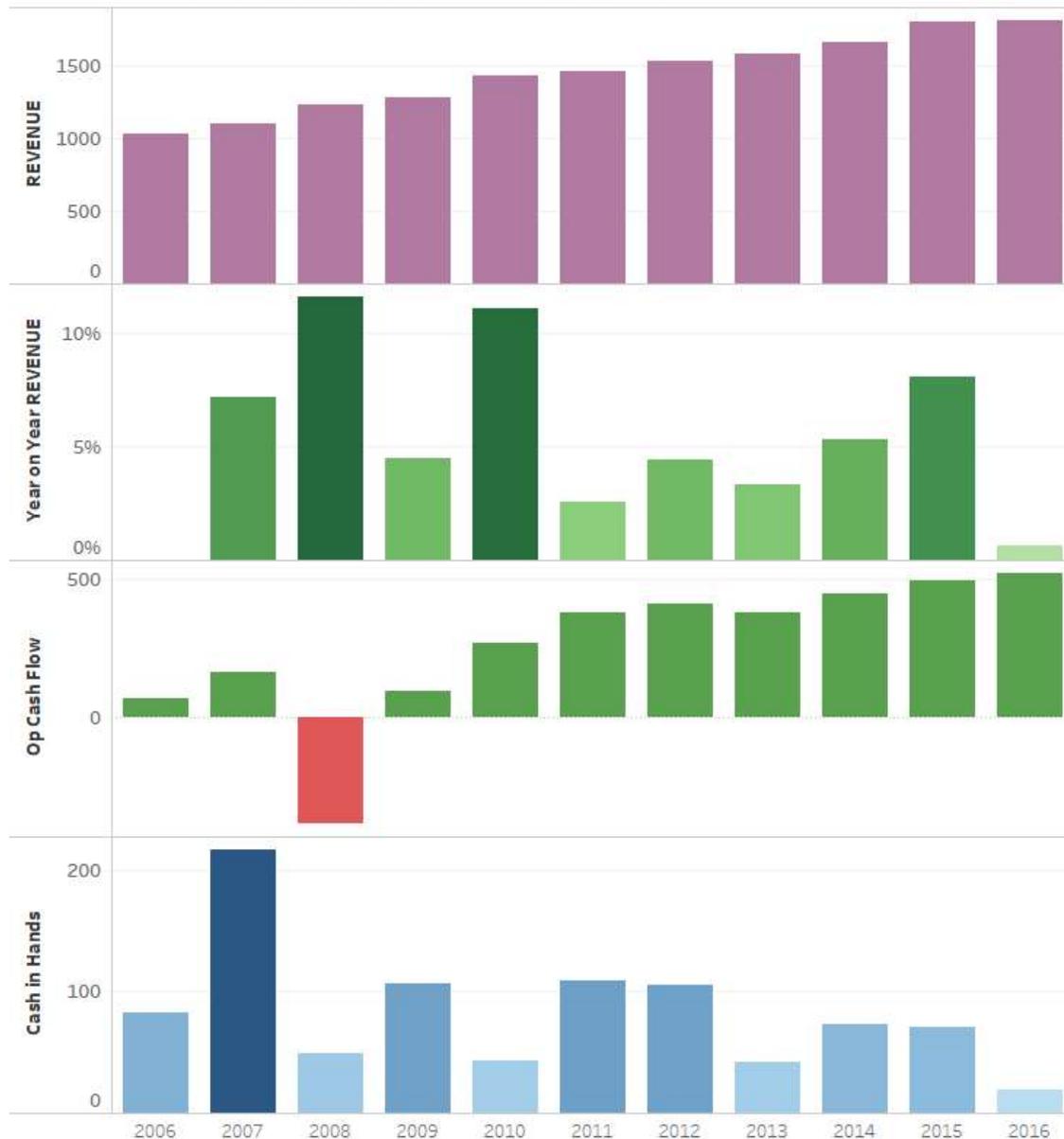

Very relevant to the situation of the EPO is the observation of cash-flows. Only once in the last 11 years was the Operating Cash Flow not positive (the effect of a ruling on income taxability of pensions). This demonstrates the capacity to conduct operations efficiently.

In the meantime, the overall cash-flow has been positive and negative, but it is noticeable that the 11-year trend is a decrease of cash and cash equivalents, from 146M€ to 19M€. This is due to the increase of negative investing cash-flows, driven by the purchase of RFPSS assets, the funding of improvement projects (IT systems and developed software are capitalized at cost) and the construction of the new Main Office building in Rijswijk.

Financial cash-flows are negative but limited to small amounts like repayments of interests on leases. There is no significant positive financial cash-flow as the EPO does not issue equity.

It is our analysis that 2017 will be a key year, as the EPO might be forced to liquidate some financial assets in order to have the cash necessary for operations. This short overview demonstrates the importance of generating operating cash-flows, in order to retain financial independence from the participating nations. Yet the trends do not make clear if the burden of investments will overpower the capacity to generate operating cash-flows.

In that perspective, we can refer to (Moore, 2015), "Zone to Win", and put the focus on the "Performance Zone" and the "Productivity Zone":

- Performance Zone: the processing of patent applications by patent examiners. An accent should be put on forecasting demand, as well as output.
- Productivity Zone: the production of systems for efficiency. The accent is on creating a different culture, and delivering the programs that improve the effectiveness of the work done in the Performance Zone.

## 3.3 STAKEHOLDERS AND ORGANIZATIONAL CLIMATE

The economic interest of the public at large is embodied in the EPC, as the granting of patents and associated monopoly is a service performed in the name of the public.

The interests of the public are of two kinds:

- The protection of the public, by enforcing unpatentable areas of activities
- The extension of public knowledge, by exchanging protection against a publication of inventions

The public is represented by the European Patent Organization (EPOrg), who has the power to modify the European Patent Convention, while the European Commission[8] leverages the use of patents in multiple stimulation programs for the European industry.

A specific part of the public is the inventors and the innovative companies. They benefit from the work done at the EPO, as if offers them the protection of revenues coming from inventions. These inventors and companies have a positive business case for innovation, through using patented technology, or through licensing to other parties who can develop products and services on top of that patented technology.

The impact of the patents on the economy is the subject of studies from the EPO Chief Economist Office[9].

---

[8] See https://ec.europa.eu/growth/industry/intellectual-property_en
[9] See https://www.epo.org/about-us/services-and-activities/chief-economist/studies.html

The IP attorneys and practitioners have physical and daily contact with the EPO. Specific web services are available to them, to track progress on their filings and expected actions. They have also an interest in searching relevant prior art, either through the tools of the EPO (EspaceNet) or through third-party services.

There is an ecosystem of companies providing services centred around the data-driven management of patent portfolios, based on data feeds from the EPO and other patent offices. Technological trends can be identified, as well as estimation of economic value of patents or applications, or automated infringement detection. The trend towards data-driven insights is clear from the customer side, and it relies mainly on the curation of incoming data, which is performed by patent offices[10].

A saliency assessment of stakeholders according to (Mitchell, et al., 1997) gives a first mapping over the relative importance of them:

- Definitive Stakeholders: EPOrg, Administrative Council
- Dependent Stakeholders: Inventors, Companies, IP legal firms. They have Legitimacy and Urgency, but lack Power on their own, and have to rely on advocacy of others. The staff at the EPO is in the same situation.
- Latent Stakeholders: the public at large, the European Commission

The organization operates in a typical Input-Output model of the corporation, where the firm is designed to meet the needs of the Customers, in that case, the innovative companies and their attorneys. It implements an Agency view (Perrow, 1986), fostering self-interested behaviours through the leadership style and the performance and reward management process.

According to the organizational context framework of (Birkinshaw & Gibson, 2004), the EPO is perceived by the public as operating under a "Country Club Context", while it advertises himself as an employer offering "High-Performance Context" to highly-skilled staff. The contradiction is a sign of the changes the EPO has to go through in order to improve its efficiency, in the face of financial uncertainty. It is only logical that the EPO is opening positions in the field of "Employer Branding", in order to fill that gap.

As it is observed when corporations use debt in order to enforce lean and productive management by the executives, the awareness of the cash situation is pushing the EPO to the High-Performance Context, the natural context of a properly working Performance Zone.

## 3.4 STRATEGY

In 2010, the newly appointed president[11] requested audits of the Finances and the IT system of the EPO. A new strategy and a roadmap were devised from the conclusions of these audits[12].

Regarding the finances, the willingness to stay self-sustained forces a strategy of generating increasing operating cash-flows, under the forecast of limited growth of revenues. The EPO has a monopoly for the grant of European Patents, while there is competition for the processing of PCT filings, between the 22 designated offices. Process Efficiency and cost containment are then the natural priorities:

---

[10] See for example www.octimine.com
[11] Benoit Batistelli, see https://en.wikipedia.org/wiki/Beno%C3%AEt_Battistelli . His mandate was renewed in June 2014 and will end in June 2018
[12] Undisclosed Internal Documents

- Improve the output of European Patents.
- Maintain the competitive position with regards to the PCT activities (EPO owns 38% of the "market")

It is not expected that fees could be significantly increased, beyond the usual inflation adjustments, and the volume of patent application is highly correlated to economic cycles in the underlying industries.

Regarding the IT systems, the main line is to consolidate the position of EPO as a leader in Search and Patent Information and maintain this position against the improvements and efforts from other patent offices, in face of a fast increase of the volume of available and relevant documentation. Furthermore, it is expected from IT systems and software to participate in efficiency improvement.

The challenge is to reconcile the dire necessities of an efficiency plan with the needs of a highly skilled knowledge-intensive workforce. Another conflict is likely to emerge when more organizational agility will be needed to implement the reengineering effort, while public international organizations have a natural tendency towards risk-aversion. We recognize these conflicts and challenges, and that only a strong and smart leadership will be able to lock the trust needed to go forward.

The culture of change is an important aspect of the Productivity Zone, this will prove to be a challenge organization-wide.

A specific focus is given on the task of search, in order to improve navigation through documentation repositories and provide as much automated search as possible, in order to reduce the time examiners spend on searching relevant documents. The vision of the operations and supply-chain problem is that an increase of the search efficiency will create room for a higher productivity, and therefore higher revenues on the same cost base.

In a typical efficiency improvement setting, project benefits will be measured in number of additional products per year throughout the office, to ensure alignment with the strategy.

We recognize the need for an office-wide Big Data Strategy that would map the office into an operations and supply-chain problem, and provide tools and support for decision-making, project design and project selection.

# 4 SEARCH FOR PRIOR ART

## 4.1 DESCRIPTION

The EPC states the merits of a patent in Art. 52:

> *European patents shall be granted for any inventions, in all fields of technology, provided that they are new, involve an inventive step and are susceptible of industrial application*[13]

Other than the field of application (further articles reduce the scope of what can be patented), the key concepts are:

- **Novelty**, as described in Art. 54 :

    *An invention shall be considered to be new if it does not form part of the state of the art.(...)*

- **Inventive Step**, as described in Art. 56 :

    *An invention shall be considered as involving an inventive step if, having regard to the state of the art, it is not obvious to a person skilled in the art.(...)*

- **Industrially Applicable**, as described in Art. 57 :

    *An invention shall be considered as susceptible of industrial application if it can be made or used in any kind of industry, including agriculture.(...)*

The search for prior art is the cornerstone of the evaluation of the merits of an invention, as it is the basis from which the novelty and the inventive step are appreciated by patent examiners.

Applicants cite prior art in their application, and argue on the merits of their invention based on the distance between the cited prior art and their creation. Patent examiners scroll through documentation repositories searching for potential other relevant documents that might invalidate the claims of the applicant. Other parties, mainly competitors of the applicant, or even the public at large, are also invited to submit their share of relevant documents and their observations on the effect these documents can have on the sustainability of a claim of novelty.

There is a part of interpretation to evaluate whether an invention is a clear disruption in the field, or only an incremental logical improvement, but the foundation of a well-informed examiner opinion is a thorough search, both compact (high precision) and relevant (high recall).

Each citation of the search report is associated with a code that indicates in what way the cited document relates to the patent application. For this study, we will consider as relevant the documents annotated as 'X', indicating that their contents are destroying the claims of novelty and inventivity of the patent application.

---

[13] EPC, Art. 52. See http://www.epo.org/law-practice/legal-texts/html/epc/2016/e/acii_i.html

## 4.2 TOOLS AND SETUPS

The tools of the Search for Prior Art consist mainly of a Boolean query interface targeting a search engine. In a first instance, after reading an application, the examiner produces his own list of important keywords and formulates a query to a search engine, or even to a few different search engines, as there are multiple disconnected documentation repositories. This is very alike to a standard Google-search, another wrapper taking care of storing retrieved documents that the examiner deemed relevant, for future citation and annotation.

## 4.3 ACADEMIC ENVIRONMENT

In 2009, the TU Wien (Austria) launched CLEF-IP[14], a yearly conference track dedicated to Information Retrieval in the context of Patent. Based on public datasets (MAREC[15]), and curated datasets (CLEF-IP[16]), this track was focusing on automated tools to retrieve relevant documents based on the text of application documents.

The task "Prior Art Candidate Search" of CLEF-IP 2011 was centred on retrieving prior art, and the organizers produced a specific dataset of applications with known relevant documents (Gold Standard). Our work will make use of the published datasets of CLEF-IP 2011 and of the Gold Standard for evaluation against the currently used tools at the EPO.

There is an ecosystem of companies providing services to IP lawyers or IP departments of established companies. Most of the services revolve around search, prior art search, linking patents to identify industry trends, evaluating the economic market value of a patents (patents are assets valued at registering costs, their market value being an estimation of future free cash flows generated by the exploitation of the monopoly on the invention), or detection of potential infringements.

## 4.4 METRICS

As an Information Retrieval problem, we can choose from the whole range of typical metrics of the discipline[17].

The publications from CLEF were a major source of literature and inspiration for this work, one of them including the development of a new recall-like metric specific to the problem of patent retrieval, which is a recall-oriented task with a high number of documents in the repositories, while only a handful of them are relevant to a specific patent application, and the ranking of results is highly relevant to the end user. This metric, the Patent Retrieval Evaluation Score (PRES (Magdy & Jones, 2010)) will be used to establish performance differences with the standard, rather than MAP[18] or Recall.

We will focus on using PRES and Recall as metrics of choice, they both are numbers ranging from 0 to 1, the higher the better. Recall will indicate which fraction of the relevant document we expect to retrieve, keeping in mind that for each Patent, there might be only a handful of relevant documents when it comes to arguing against novelty and inventivity.

---

[14] See http://www.ifs.tuwien.ac.at/~clef-ip/
[15] See http://www.ifs.tuwien.ac.at/imp/marec.shtml
[16] See http://www.ifs.tuwien.ac.at/~clef-ip/download-central.shtml
[17] https://en.wikipedia.org/wiki/Information_retrieval
[18] Mean Average Precision

Nonetheless, a critical review of the specificities of the task will help to focus on only a handful of them. Typically, there are up to 5 relevant documents per search. The search will be done against a corpus of a few millions documents and will return up to 1000 results.

In this setting, Recall will have only 6 potential values (0.0, 0.2, 0.4, 0.6, 0.8, 1.0) depending on the number of relevant documents in the search results. Precision will range from 0.0 to 0.005, which makes it impractical to use. PRES will vary from 0.0 to 1.0, depending on the number of relevant documents retrieved and their ranking.

We had to correct the computation of the sum of ranks for PRES, the original was:

$$PRES = 1 - \frac{\frac{\sum r_i}{n} - \frac{n+1}{2}}{N_{max}}$$

$$\sum r_i = \sum_{i=1}^{nR} r_i + nR(N_{max} + n) - \frac{nR(nR-1)}{2}$$

n is the number of relevant document, R is the recall, and $r_i$ is the rank at which each relevant document is found.

The assumption is that for each relevant document found after the rank $N_{max}$, and for each non found relevant document, the calculation is done as-if the document was found at rank $N_{max} + n, N_{max} + n - 1, etc …$

With this example in mind, using $N_{max} = 100$:

| Relevant Document | Found at rank | Using rank for PRES@100 |
|---|---|---|
| D1 | 97 | 97 |
| D2 | 85 | 85 |
| D3 | 87 | 87 |
| D4 | 625 | 104 |

$n = 4$ and $Recall@100 = 0.75$

The document D4 was found after rank 100, so it should be considered as found at rank 104, and then from the formula above, we get:

$$\sum r_i = (97 + 85 + 87) + 3*(100+4) - \frac{3*(3-1)}{2} = 578 \text{ and then } \boldsymbol{PRES = -0.42}$$

We have to correct this formula to make the correct computation:

$$\sum r_i = \sum_{i=1}^{nR} r_i + \sum_{i=nR+1}^{n} (N_{max} + n - (i - nR - 1))$$

With this in mind, now we compute

$\sum r_i = (97 + 85 + 87) + (104) = 373$ and then $\boldsymbol{PRES = 0.0925}$ which is correct, as the 3 documents found before rank 100 were found at very high rank, so we are close to the worse-case scenario for PRES.

## 4.5 STATE OF THE ART

Early this year, an overview of published work on patent retrieval (Shalaby & Zadrozny, 2017) focused on comparing various systems and their performance with regards to the CLEF-IP Patent Retrieval task. It mentions the many challenges of the patent retrieval task: "*patents are multi-page, multi-modal, multi-language, semi-structured, and metadata rich documents*". Multi-modal means that a patent contains text as well as illustrative images, and text about the images.

The current generation of tools rely on generating queries for search engines, based on the full-text of a patent application. The Query Re-Formulation (QRE) approach consists in identifying potential keywords in the document (keyword-based QRE), and potentially expanding with synonyms (semantic-based QRE), the more advanced tools making use of user feedback to refine the queries and results (interactive QRE).

Queries are modified in 2 ways:

- Reduction (QR): only a significant subset of words is retained as keywords, from the original full text query
- Expansion (QE): words out of the original full text are added to the query, for example synonyms of the most important concepts

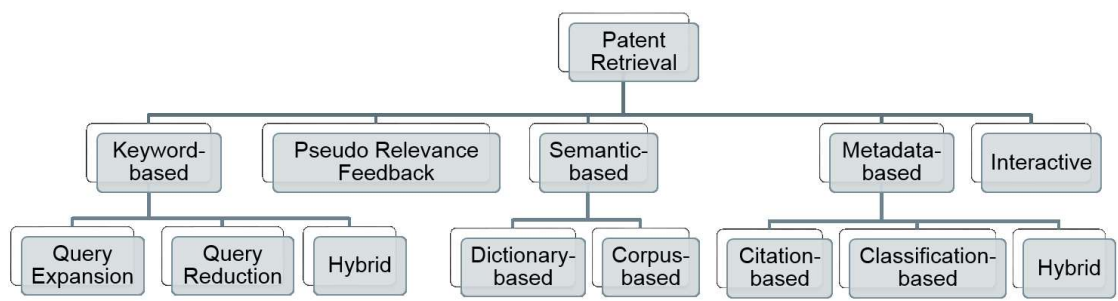

Mihai Lupu (Lupu & Hanbury, 2013) considered the landscape as well and observed the trend towards Technologically Assisted Review, where the system not only searches for relevant prior art but also provides context and description of the invention technological environment to the user.

Highest level of performance with PRES@1000 (considering the first 1000 search results for finding relevant documents) lies between 0.4 and 0.5, although we'll discuss later about the use of a more restrictive maximum rank.

Keyword-Based systems range in complexity from straightforward query formulation to elaborate multi-stage systems such as Patatras (Lopez & Romary, 2009), with QR stage being largely based on word frequency analysis methods inspired by TF-IDF, as for example in (Mase, et al., 2005).

Semantic-based methods are elaborated around a QE stage where synonyms and hyponyms[19] are found by ways of dictionaries built ad-hoc from the corpus. Although the concept is promising, no significant effect is seen on the performance, as measured by PRES@1000.

A significant assumption is that each method is tested against a gold standard that encompasses all 8 domains[20] of patents, whereas exploratory data analysis would show structural differences in the full text of patent documents depending on their domain or CPC sub-class. Later we'll discuss as well the opportunity to segregate and have methods that perform better in certain domains than others.

We observe only few efforts in the field of isolating meaningful keywords from the text based on a more semantic approach, in line with the specificities of the available full text. The very specific idiom used is mainly based on breaking down the invention in sub-parts and providing details on these sub-parts, and going more in depth on each detail. We identify a potential for an analysis of the text based on this "split-and-detail" construction, as researched by (Suzuki & Takatsuka, 2016) in a setting to describe the novelty of the invention at hand. Their research was using "cue words" to infer when a sentence chunk was detailing another one. As the target was Japanese documents, it would prove a challenging project to gather a proper list of cue words and associated sentence constructions in English, French and German. We will propose another way of extracting the detailing structure of the text.

## 4.6  EPO ENVIRONMENT

The current method used at the EPO for semi-automated search is based on Lucene, and can offer comparisons to 2 baselines:

- Using keywords extracted from the full text of patent document with an optimized TF-IDF
- Making use of native More Like This[21] (MLT), with the full text as input

The MLT baseline is the most flexible, as it can be tuned to use only the text from claims, and to search against claims only.

In this project, we used Python as a software language, and made use of python packages *nltk*, *gensim*, *pandas*, *numpy*, *sklearn*. Furthermore, we also made extensive use of Stanford NLP Tools (Manning, et al., 2014) and developed a Java annotator for the CoreNLP Server.

---

[19] "Purple" is the hyponym of "Color" (who is the hypernym), while "Red" and "Blue" are co-hyponyms
[20] See classification in 2.3
[21] See https://lucene.apache.org/solr/guide/6_6/morelikethis.html

# 5 THE TEXT FROM PATENT DOCUMENTS

## 5.1 MAKING A COMPARISON

The literature at large considers the text from patent documents to be highly specific, and indeed patent practitioners are a highly skilled workforce. We provide some elements of context in order to show how this complexity is built.

The comparison corpus is a part of the Brown corpus named "Learned", a collection of 80 books from various academic fields. This corpus is well suited for a comparison with patent texts as it describes many scientific experiments, theories, in basic and applied sciences, including physics, astronomy as well as political sciences or philosophy. It is legitimate to see this as an equivalent for the description of technical inventions, as done in patent texts.

The main drive for this short study is to get an evaluation of which NLP tools and processes might apply to claim texts.

## 5.2 COMPARE BROWN AND PATENT CLAIMS

The plot[22] of the frequency distribution (Figure 1) of the number of words in sentences shows already some significant differences between the two collections of text. Claims are made of longer sentences in number of words.

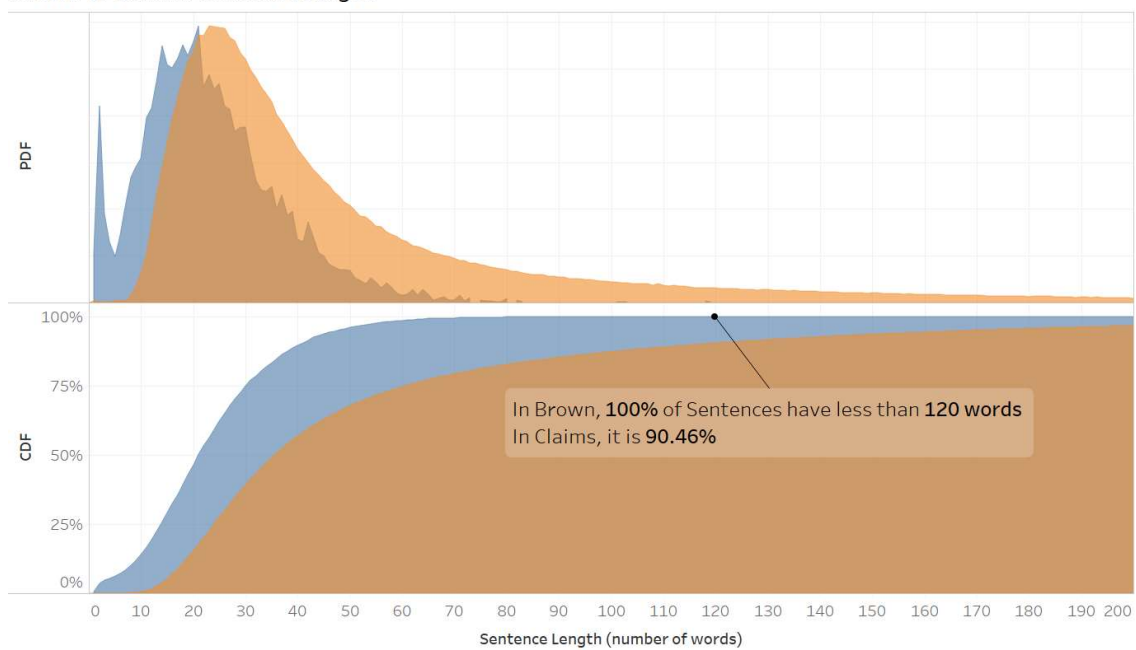

*Figure 1 - Sentence Length*

The orange represents the data from patent claims.

The distribution of sentence length in patent claims is characterized by heavy tails, and a higher value for mode and median compared to Brown Learned corpus. In our further analysis of patent

---

[22] Visualizations done with Tableau®, under Academic Licence. See www.tableau.com

claims, the processing will have to deal with long sentences, this can be an issue for standard NLP tools such as parsers

The Stanford CoreNLP default parser (PCFG) is using quadratic space and cubic time with sentence length[23], it is therefore preferable to choose the shift-reduce constituency parser. It is based on a Neural Network, and produces quick results.

We also use the Flesch-Kincaid Reading Ease[24] test to show how patent litterature is difficult to read. We computed the Reading Ease score for sentences in claims or descriptions in Figure 2.

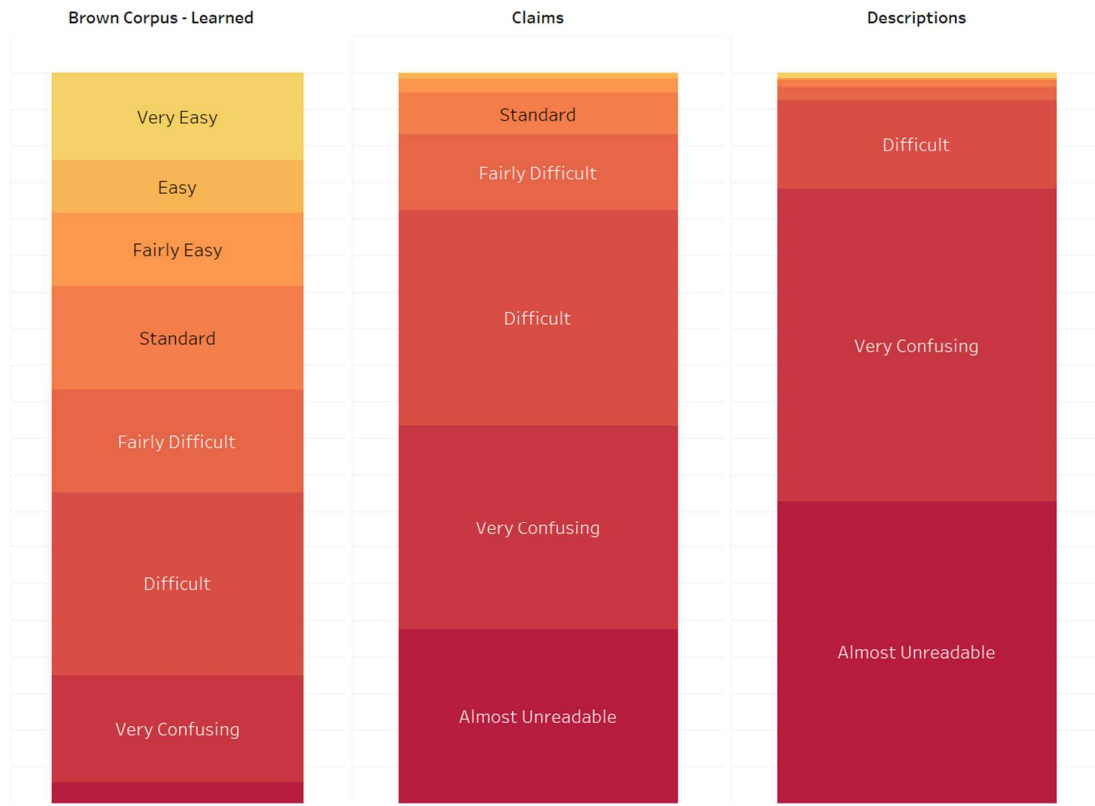

*Figure 2 - Flesch-Kincaid Readability*

The differences between standard academic literature and patent literature are made obvious here, emphasizing the need for specific skills. On the other hand, it also means that sentences will be harder to parse than usual.

For our study, this means we should always keep an eye on the quality of the work done by the used NLP tools, especially the parsing and POS-tagging, as the texts we are going to use cannot be considered as similar to those used for training these algorithms.

---

[23] https://stanfordnlp.github.io/CoreNLP/memory-time.html
[24] https://en.wikipedia.org/wiki/Flesch%E2%80%93Kincaid_readability_tests

## 5.3 EXTRACTING KEYWORDS

An efficient QR method will rank important words from the text of the claimset, and put forward the N first words as being keywords. In a typical setting, 100 keywords are extracted from the text, and further optimization shows that recall performances of searches increase with the number of keywords used for the search, from 10 up to 70.

In our case, the algorithm achieved best results for 100 keywords, which means all the extracted keywords were useful, and in the order in which they appear, they continued to add (even if very little) information.

The challenge is as well to be able to extract N unique words out of the original text, when typically the claimset of a patent has a limited number of words.

For example, extracting 100 keywords from any claimset leads to being forced to use more than 20% of the words of the text as keywords in 30% of the cases, as shown in Figure 3. Obviously, this will create noisy keywords.

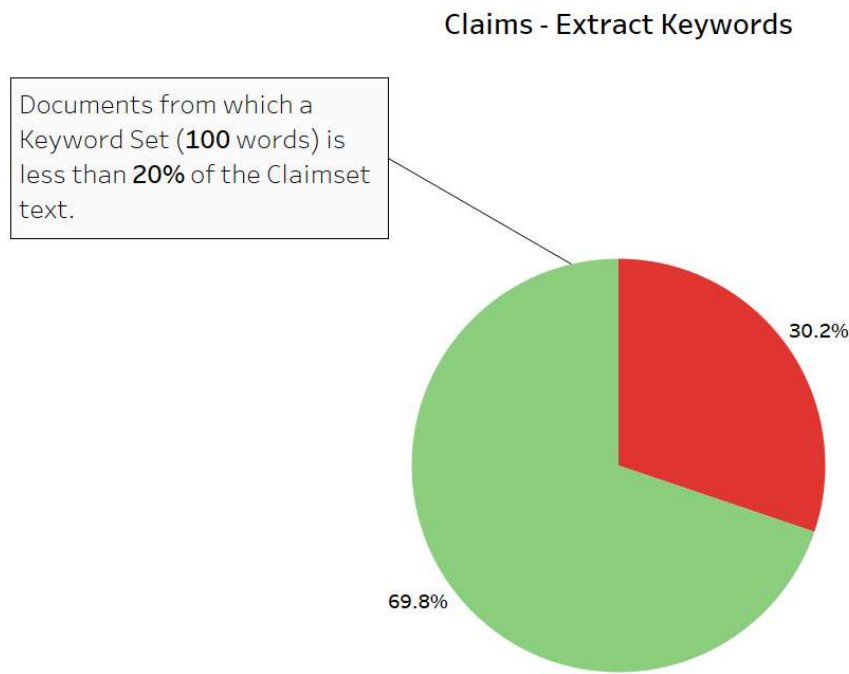

*Figure 3 - Keyword weight in Claim Set*

Nonetheless, in a search setting, it is relevant to keep as many keywords as possible, and later observe how performance is affected by the number of keywords.

In an exploratory data analysis, or a clustering setting, we would decide to reduce the list of keywords to 10 or 20.

## 5.4 STRUCTURE OF THE TEXT OF PATENT CLAIM SET

The claims are the expression of the protection that the applicant is requesting, as in Article 84 EPC:

> *The claims shall define the matter for which protection is sought. They shall be clear and concise and be supported by the description.*

The Rule 43 EPC gives more details on the Form and Content of Claims.

As an example, we can read in patent document "EP 1221372 A2"[25], an invention relative to "Integrated programmable fire pulse generator for inkjet printhead assembly", the claims read as follows:

Claim 1 :

> *An inkjet printhead (40) comprising: nozzles (13); firing resisters (48); and fire pulse generator circuitry (100/200) responsive to a start fire signal to generate a plurality of fire signals, each having a series of fire pulses, by controlling the initiation and duration of the fire pulses, wherein the fire pulses control timing and activation of electrical current through the firing resisters to thereby control ejection of ink drops from the nozzles.*

Claim 2 :

> *The inkjet printhead of claim 1 wherein the fire pulse generator circuitry comprises: pulse width registers (110) for holding pulse width values, wherein the duration of the fire pulses is based on the pulse width values.*

We can already see here a few features of claims :

- They can reference each other. This is referenced to as the "claim tree" (Dong, 2014), although each claim can have multiple parents and multiple children. In this paper, we'll consider the depth of a claim as being the number of times we should navigate up in the claim hierarchy to reach a claim without parents.
- Claims provide details about the invention, each sentence fragment providing details about its previous sentence fragment. We will base our keyword extraction on this specific feature
- The language is unusual, with few verbs

Claims have their own idioms, the most famous one being the use of the word "said", which is almost always a relative adjective, as opposed to a conjugation of "to say". NLP tools that were trained on generic literature will likely misclassify "said" as a verb, when doing the POS-tagging, which could lead to potential issues further down the processing pipeline, as a parser will then be forced to accept ill-formed parsing, in order to accommodate this verb.

This was observed in (Hu, et al., 2005), we have re-used their method and dataset to correct for that effect. In effect, we implemented and trained a Support Vector Classifier, based on annotated fragments of claim, with the proper POS-tagging for words that were tagged as VERB by the initial POS-tagging.

---

[25] https://encrypted.google.com/patents/EP1221372A2?cl=en

The original dataset (Campaign 4: 124,000 data points) is split 80-20 between a training set and a test set. We make use of Python package *sklearn* to instantiate a Linear SVC[26] problem. The Linear SVC problem makes use of a linear kernel (dot product). Each datapoint is a 900 dimensional vector, labelled with the correct POS tagging.

For example, Figure 4 shows the initial parsing of a sentence :

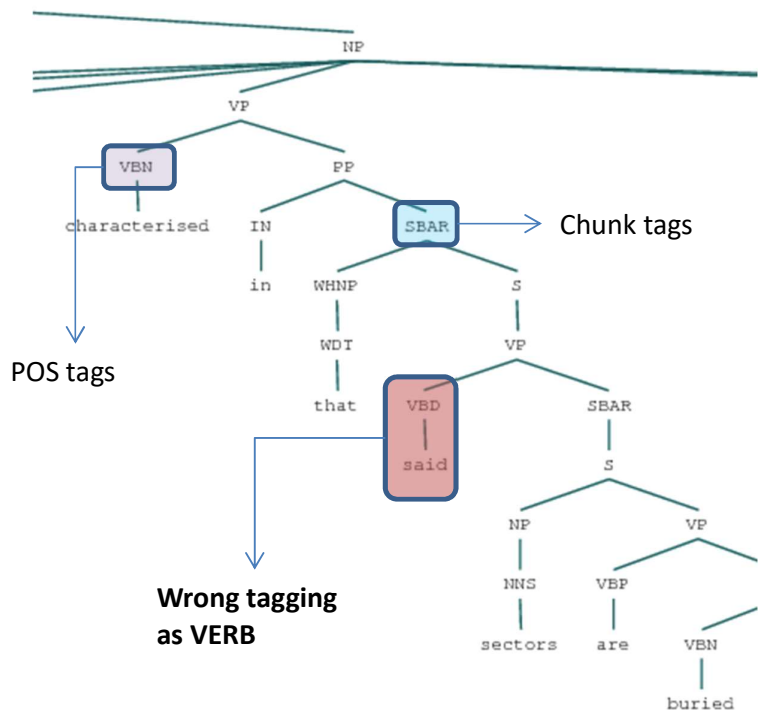

*Figure 4 - Initial Parsing*

We trained an SVC to correct the POS tagging, the input for that classifier being the embeddings of the Trigram[27] (in our case "that said sectors") where the VERB is the middle word. The embeddings are taken from a published Google model (GoogleNews), we advise in the future to re-use an embedding model based on Patent Literature, to account for the specific use of words in the context of a Patent as opposed to its use in the more general context. The SVC features a Linear Kernel, after training with the published datasets from (Hu, et al., 2005), the SVC has a training Mean Accuracy of 0.9.

---

[26] http://scikit-learn.org/stable/modules/generated/sklearn.svm.LinearSVC.html
[27] 3 words

The sentence can be re-parsed, as displayed in Figure 5.

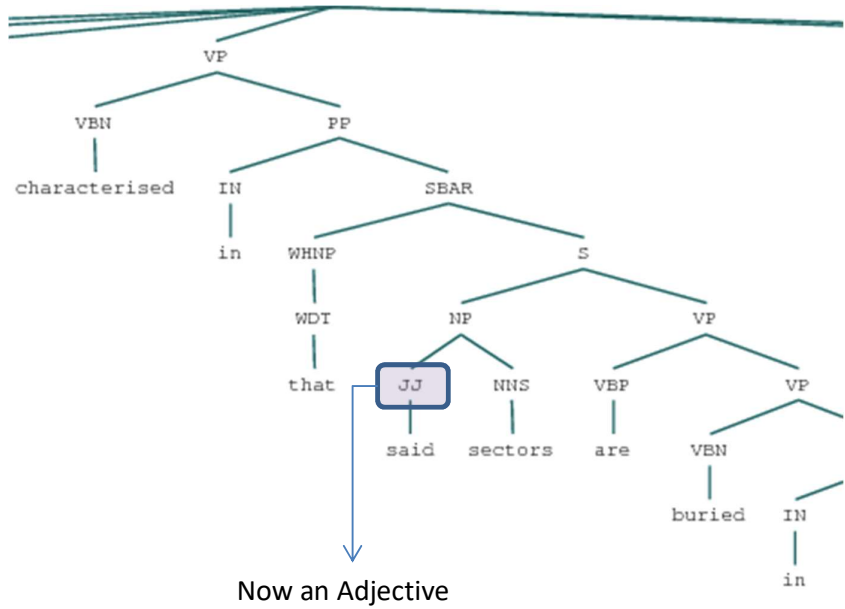

Now an Adjective

*Figure 5 - Corrected Parsing*

# 6 MEANINGFUL KEYWORD EXTRACTION

## 6.1 THE SPECIALIZATION TREE

As described above, we will take advantage of the usual way of writing claim statements, which goes gradually into the details of the invention. Each sentence fragment specializes another one, according to a structure with 2 relations:

- Aggregation: typically when enumerating components. See the Claim 1 of the example patent document.
- Specialization: when a technical detail is given. We can illustrate the formation of a tree based on the Specialization

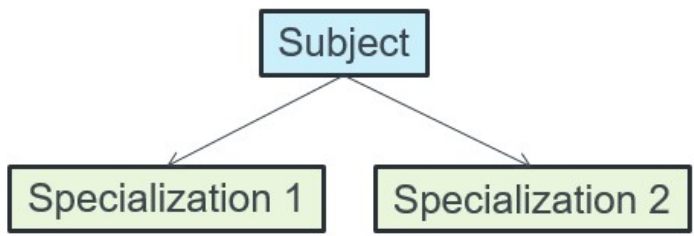

Figure 6 - Creating a Specialization Tree

For example, if we use the sentence (EP-1748300-A1, Claim 37):

> *Method according to one or more of the preceding claims 25 to 36, characterized in that initial iteration steps for determining compensation dipoles by means of quadratic or linear programming can provide in combination a modification for each subsequent iteration step consisting in a reduction of constraints such that the partial solution converges progressively towards a solution that is considered an optimum one.*

We obtain the following Specialization Tree (Figure 7):

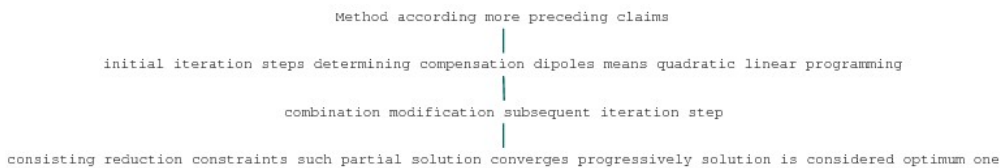

Figure 7 - Example of Specialization Tree

Each claim is considered as one-sentence text. It happens that standard tokenizers split claims into more than one sentence, although we can prove that this is only due to misunderstanding the usage of ".", as in numbering ("1.", "2."…), or being confused by complex chemical formulas.

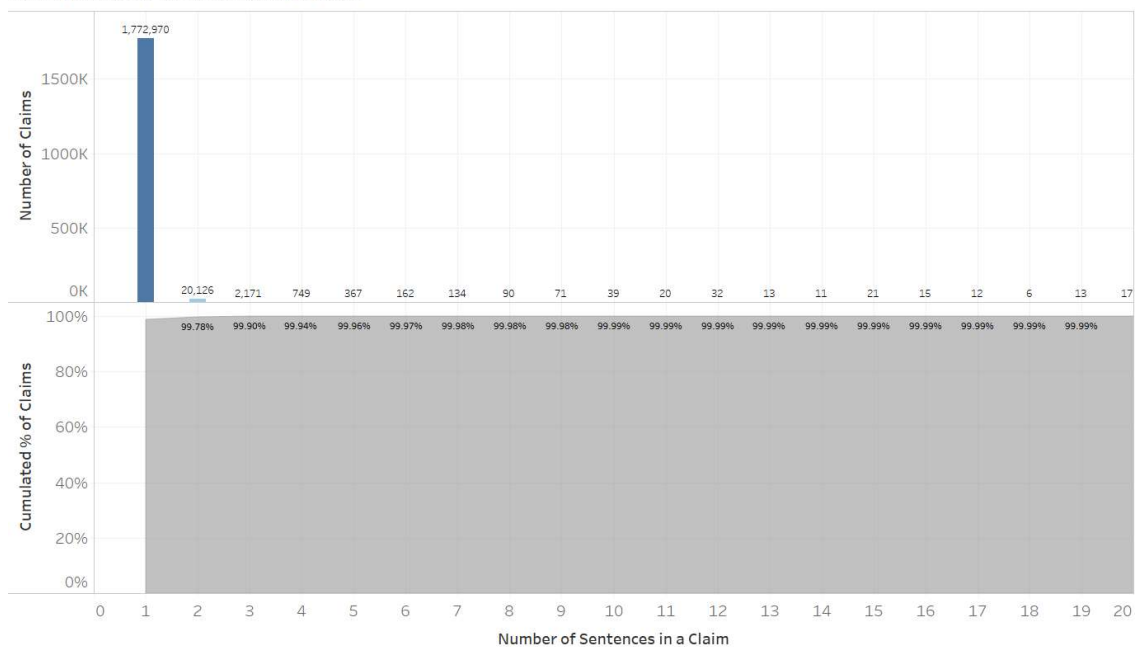

Figure 8 - Distribution of Number of Sentences per Claim

We observe that only 0.22% of claims get split into more than 1 sentence. This confirms our decision to continue this way and ignore any sentence splitting within individual claims.

The Specialization Tree of one claim is generated based on the syntactic parsing of the claim sentence. Based on our knowledge that the generic parsing tool incorrectly tags words as verbs, we'll introduce a correction, based on the POS-tagging correction, as seen in (Hu, et al., 2005).

## 6.2 THE PIPELINE

We come up with the following end-to-end process in order to process the claims and extract the keywords.

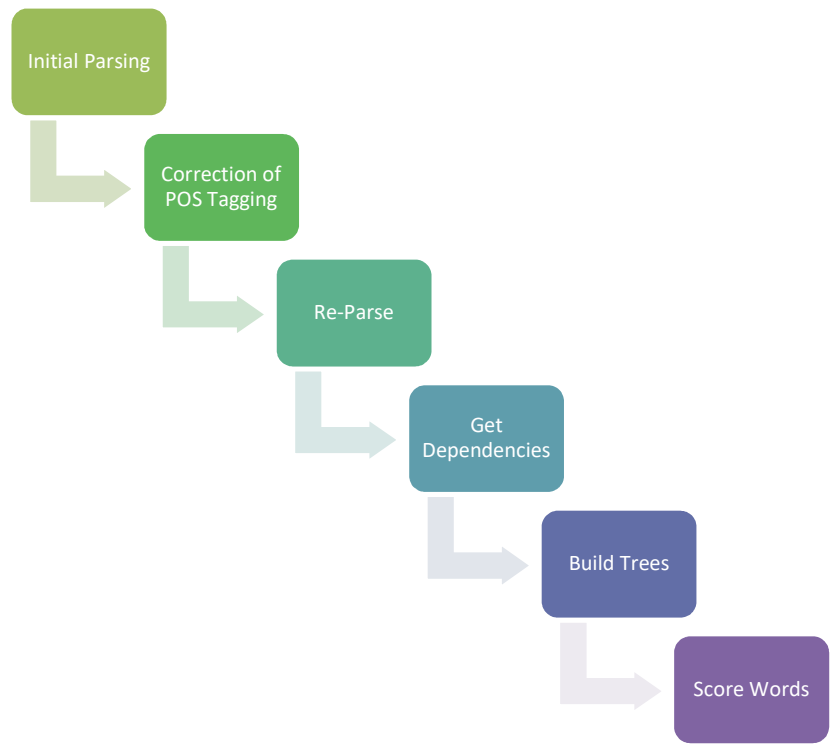

*Figure 9 - Processing Pipeline*

The parsing is done by a Stanford CoreNLP server, the POS tagging correction is handled by another developed web service that delivers the predicted correct POS tag for each VERB. The Re-Parse is done as well by a Stanford CoreNLP server, with an ad-hoc written Annotator. The Specialization Trees are built, and a scoring function is applied to the words from the text, leading to a ranking and an extraction of N most important words.

The original dataset is the english section of the Topic dataset from CLEF-IP 2011, consisting of 1,300 documents. Another bigger dataset covering 130,000 documents was also used to make sure of the scalability of the solution.

The CoreNLP server was the proper setting for performance, handling sentences up to 1000 words, where the standalone version of the Stanford Parser was running out of memory with 300 words.

The processing is done offline by batch, the Topic collection would take up to 4 hours to be processed, while the larger collection would be processed over a complete week-end. The resources available at the EPO were sufficient to accommodate the computation.

## 6.3 SCORING METHOD

We consider one Patent Document, and its claim set. Each claim generated a specialization tree. We start by identifying the location of words within the trees.

$$\forall w, N(w) = \{nodes\ n_i\ where\ w \in n_i\}$$

$$P(w) = \{(nd(n_i), nh(n_i), cd(n_i)), n_i \in N(w)\}$$

For each word, for each node where it appears, we retain the following information:

- Node Depth (nd): depth of the node within the specialization tree. The depth of the root node is 0, the depth of a node is the depth of its parent plus 1
- Node Height (nh): the height of the node within the specialization tree[28].
- Claim Depth (cd): the depth of the claim within the claim tree

The words are stemmed by an instance of the Porter Stemmer, and for each stem we retain the word with highest number of occurrences:

| Word     | Stem  | Occurrences |
|----------|-------|-------------|
| Curated  | Curat | 12          |
| Curation | Curat | 8           |

In that example, we will associate the stem "Curat" with the word "Curated". We'll run the scoring method on the stems, in order to refocus the variations around a word (plural form, conjugation of verbs, …).

$$\forall stem\ s, P(s) = \bigcup_{stem(w)=s} P(w)$$

The score for each stem in algorithm named "Juju-06" is:

$$score(s) = c * \sum_{P(s)} e^{\alpha*Max(nd)+\beta*Max(cd)}, \alpha = 1, \beta = 2$$

Or in algorithm named "Juju-05":

$$score(s) = c * \sum_{P(s)} e^{\alpha*\frac{nd}{nd+nh-}+\beta*cd}, \alpha = 1, \beta = 1$$

The main idea is to favour the words that appear deep in a specialization tree, and also deep in the claim tree. The further in depth a word appears, the more likely it is to be about a significant detail of the invention.

Where $c$ is a normalizing factor so that:

$$\sum_s score(s) = 1$$

---

[28] See the documentation of NLTK Tree Height() : http://www.nltk.org/api/nltk.html#nltk.tree.Tree.height

## 6.4 TESTING RESULTS

The Patent Documents from the Topic collection of CLEF-IP 2011 are a part of a Gold Standard, where the list of relevant documents was established for each Patent Document. The EPO has a setting dedicated to testing retrieval performance of multiple algorithms based on this Gold Standard. The Gold Standard encompasses Patent Documents in all of the 3 official languages of the EPO: English, French and German. We will focus on English documents only.

Based on the provided detailed results, we are able to compare the Recall and PRES at 10, 50, 100, 500 and 1,000.

The list of tested algorithms includes :

- Derivations based on the specialization trees, with different scoring methods, and also with or without POS-tag correction
- TF-IDF from EPO, extracting keywords from the full text (Title, Abstract, Description, Claim) of a patent document
- MLT : More Like This, using a provided text to find other relevant documents

Furthermore, the algorithms can be tested in different configurations:

- Extracting keywords from full text or from claims only
- Searching against the corpus of full texts or the corpus of claims only
- Using or not the word scores as an indication for boosting the query keywords, by applying different weighs to the keywords

This creates a large number of possibilities. In order to restrain the field of study, we establish that we want to compare the performance of algorithms that extract keywords from claims and focus the search on using only the collection of claims. In this setting, the importance of extracting meaningful keywords from claims will show.

Furthermore, a business analysis[29] would show that a patent examiner would review approximately 400 documents before writing an opinion on the novelty of a patent application. Since the search in patent literature is a part of the complete search process, we suggest that Recall @100 and PRES @100 should be the metrics of choice.

An interactive version of the results is available as a Tableau Public visualization at address https://public.tableau.com/profile/julien.rossi#!/vizhome/EPO_Metrics/RECALLBestSummary . These visualizations will stay published at this address in the foreseeable future, and can be consulted to get the bigger picture behind the results.

The results are presented visually hereafter in Figures 10 to 13. The reference algorithm (MLT) is painted orange, the statistical significance of the difference is indicated by the t-score of that difference under the null hypothesis that both systems produced the same performance. We consider a significance level of 5%. The colorcoding (Green, Grey, Red) is self-explaining:

- Green indicates a system with a significant and better performance than the reference
- Red indicates a system with a significant and worse performance than the reference
- Grey indicates no significant difference

---

[29] Internal EPO documentation, undisclosed

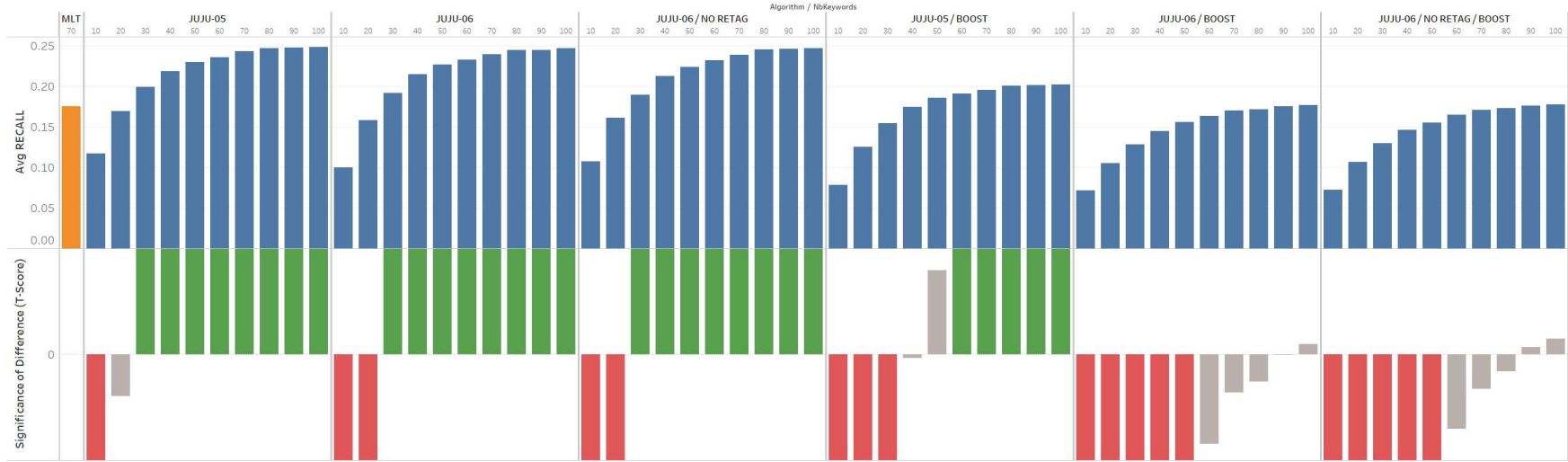

*Figure 10 - Complete Leaderboard for Recall@100*

Figure 10 shows that compared to MLT, all Juju algorithms generate higher performances that are statistically significant, even when using as little as 30 keywords. We also see that the performance of each algorithm increases with the number of keywords. Only one "boosted" algorithm generates a higher performance.

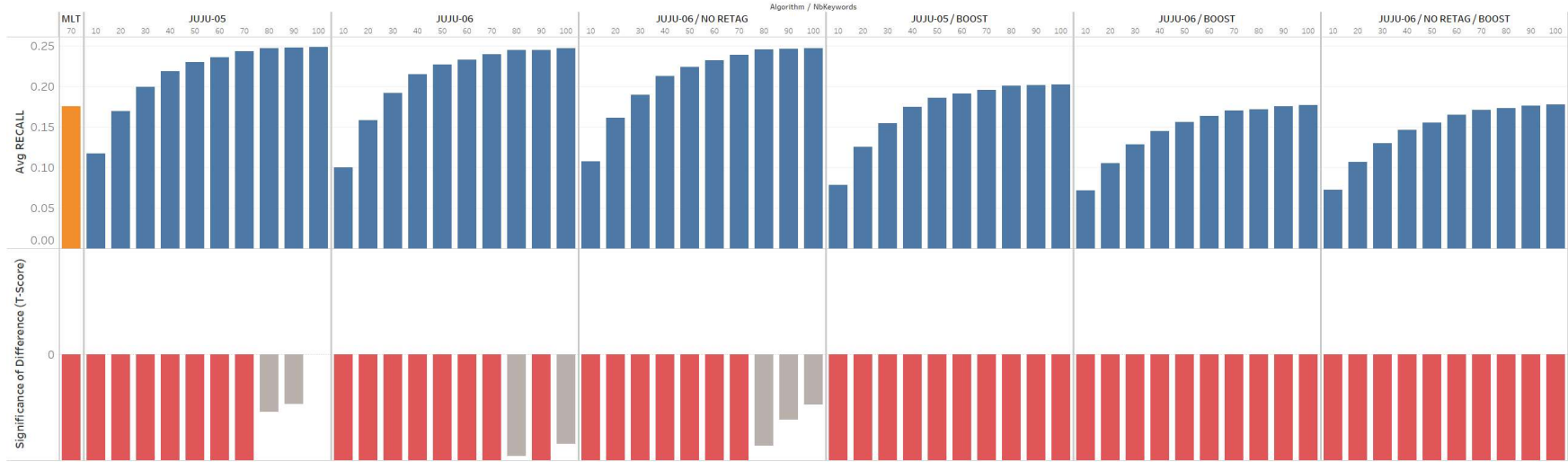

*Figure 11 - Complete Leaderboard for Recall@100*

In Figure 11, we can observe how the performance of algorithms compare with the best performer. We can see that the lower performance observed when using 80, 90 keywords is usually not statistically significant. The difference in performance between the boosted algorithms and the non-boosted algorithms is significant.

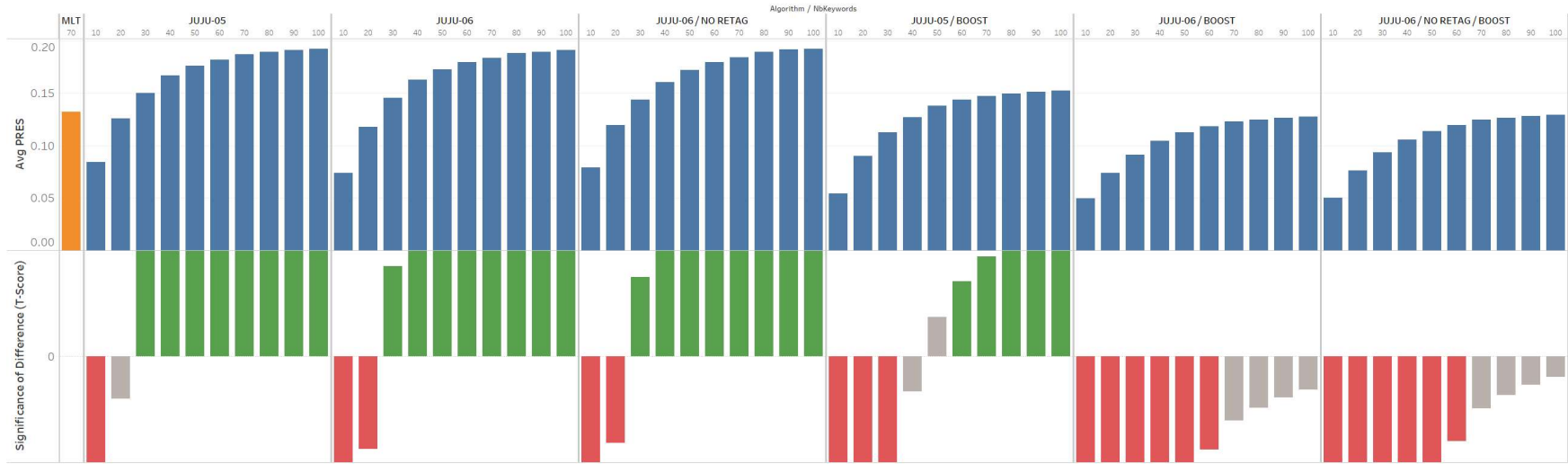

*Figure 12 - Complete Leaderboard for PRES@100*

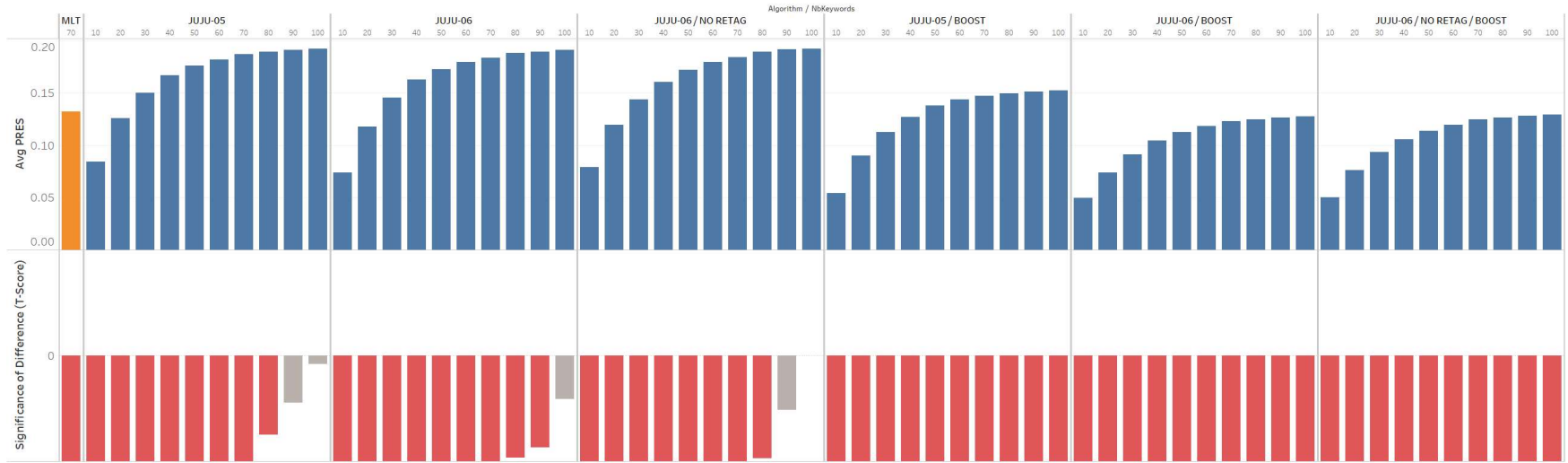

*Figure 13 - Complete Leaderboard for PRES@100*

We can see that using more keywords consistently leads to improved results, both in Recall and PRES. Although we also make note that above 70 keywords, the difference in performance is not statistically significant. So we'll consider from now on that the best performance is achieved with 100 keywords, and leave the choice to use less keyword to future users.

We observe as well that in any setting, the performance of the boosted keywords is significantly less than the best performance. We also report that only in a minority of occurrences, the boosted keywords outperformed the baseline.

We also observe from the results that a search with 30 keywords from our algorithm performs at the same level as a search by MLT, which is using the full text of the claims.

In the following figures (Figure 14, Figure 15 and Figure 16, all below), we will present a summary with only the best results achieved per algorithm, in order to get more clarity on these results.

The first line indicates the average value of the Recall@100 (or PRES@100), while the other lines contain color-coded indications:

- The statistical significance of the difference of averages, in comparison to the highest average. A full bar is always drawn, and the color reflects the statistical significance of the difference.
- The features of each algorithm: its family, if it used the POS-tagging correction, if it used Boosted Keywords or not...

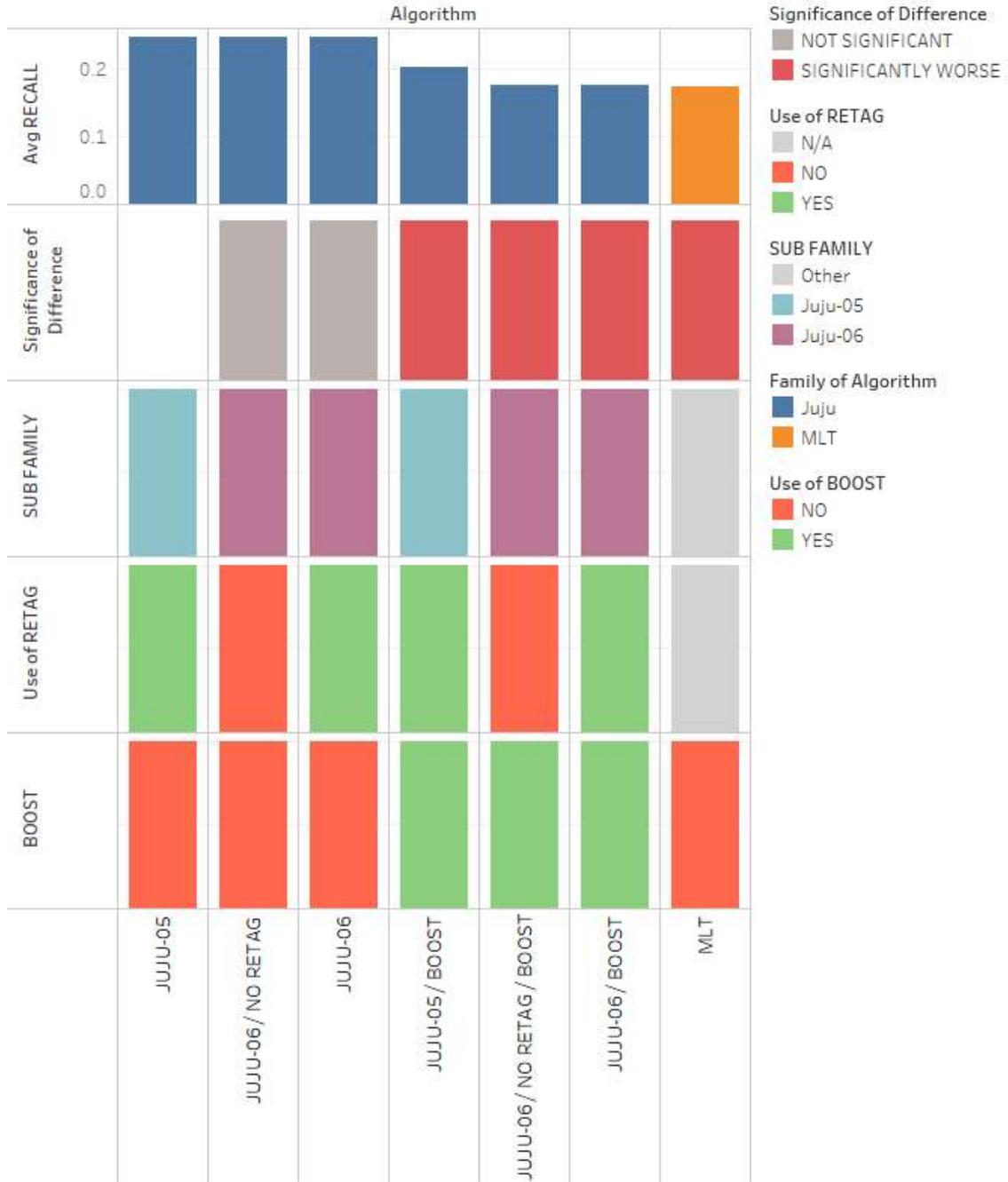

Figure 14 - Recall@100

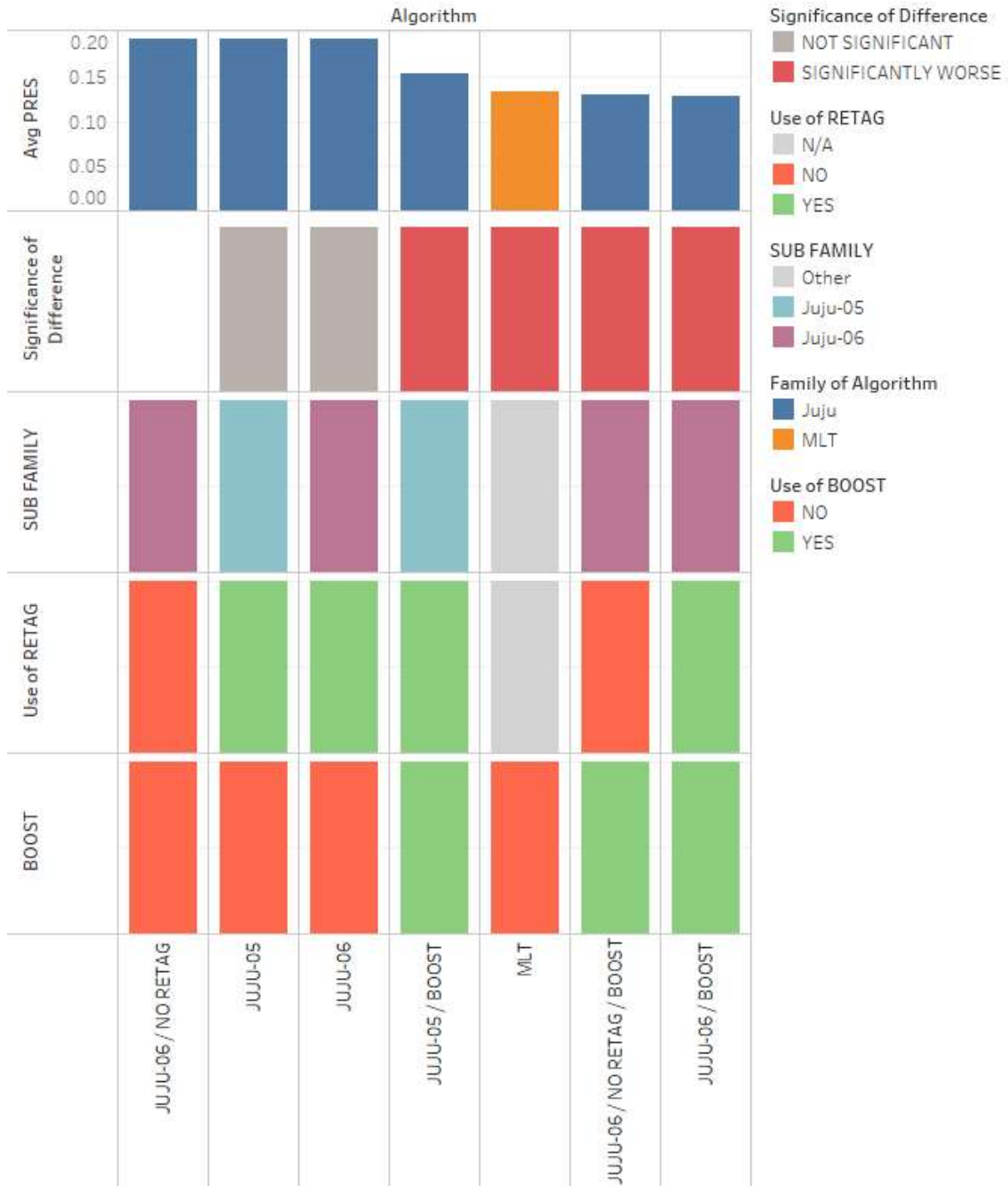

*Figure 15 - PRES@100*

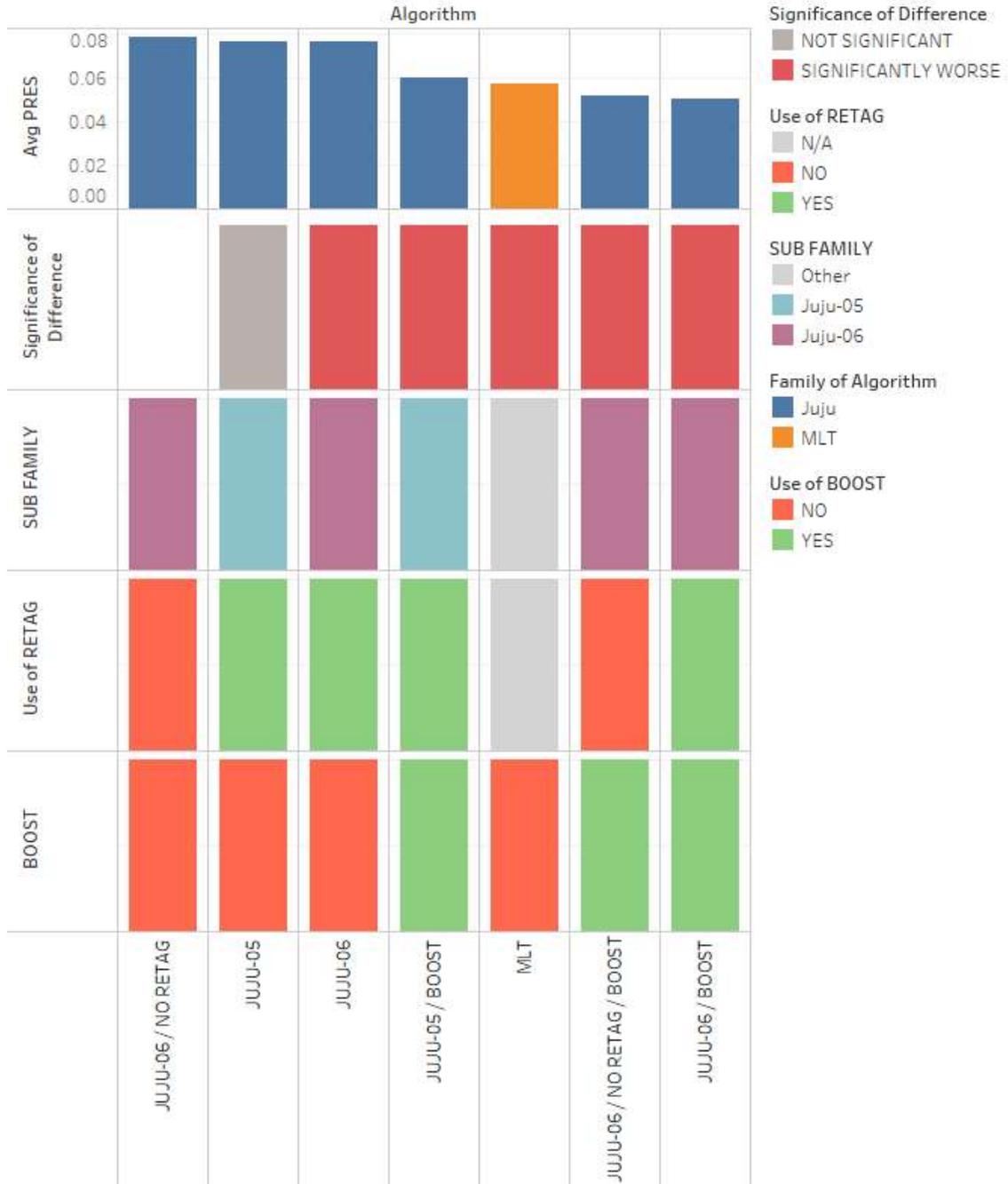

Figure 16 - PRES@10

For the 2 relevant metrics, Recall@100 and PRES@100, the keywords extracted from the claims with the Specialization Tree method (algorithm "Juju") yielded better search results when targeting the collections of claims than the only EPO comparable algorithm "More Like This".

The difference is statistically significant, as evidenced by using 2 different methods from (Smucker, et al., 2007) : the Student's t-test and Randomization test.

From the Recall@100 ranking board, we can see that for the newly developed algorithm, the 3 different flavours offer performance differences that are not statistically significant. The same algorithms with the use of BOOST, which means that the word scores are used as a Boosting factor for the search query keywords, have significantly worse performance. We can already conclude that the current scoring method does not produce a relevant boosting indicator.

The use of a corrected POS tagging does not seem to have an influence on the performance. We infer that the incorrect parsing does not affect much the recognition of the specialization in the text, and therefore down the line the scoring of individual words, as they appear at their proper place in the specialization tree. However, we consider this an important step, and will continue using it, in the light of future developments based on dependency parsing, for example.

Furthermore, we suggest as well the two following visualizations of the performance of the algorithms. The colour coding indicates the achieved value of PRES@100 for different topics of the dataset. It shows the similar behaviour of all three JUJU algorithms. The less Blue there is, the more performant the algorithm (Blue indicates no relevant document found), the more Red the better (Red indicates high rates of relevant documents found, and listed early in the search results).

The split per patent domain makes visible the difference per domain. It appears that domain C sees the best results, while G and H are the worse. This can be seen by the relative amount of Blue in the picture for one domain.

These representations include the following baseline algorithms, that are currently in production systems of the EPO:

- **MLT All/All** is the MLT algorithm using the full text of a patent, and searching for correspondences in the full text of patents in the corpus.
- **TF-IDF** extracts keywords from the full text of a patent and searches against the full text of patents in the corpus.

For a specific patent document, the different results obtained for that document can be seen by navigating vertically. Each column represents one document.

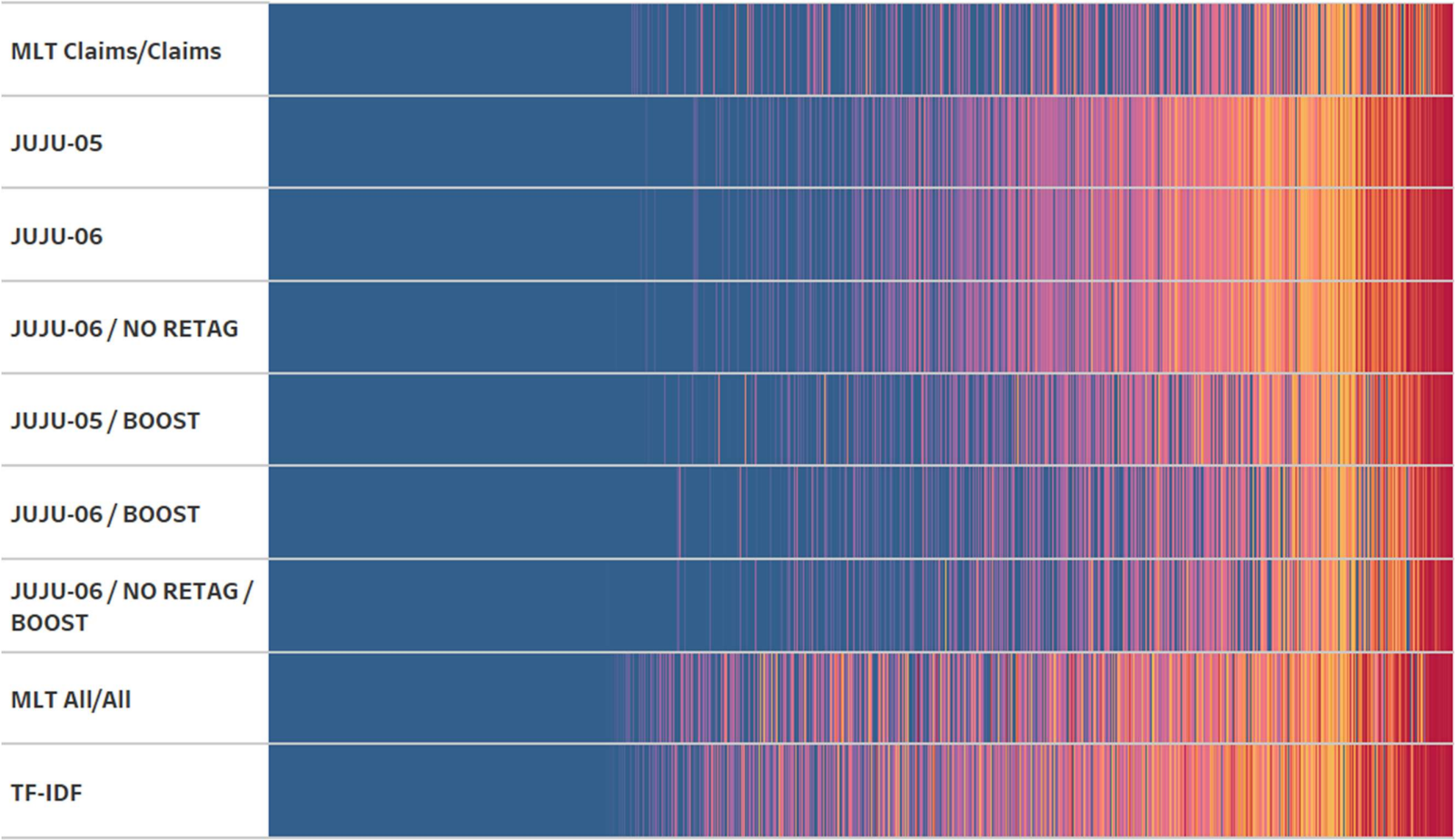

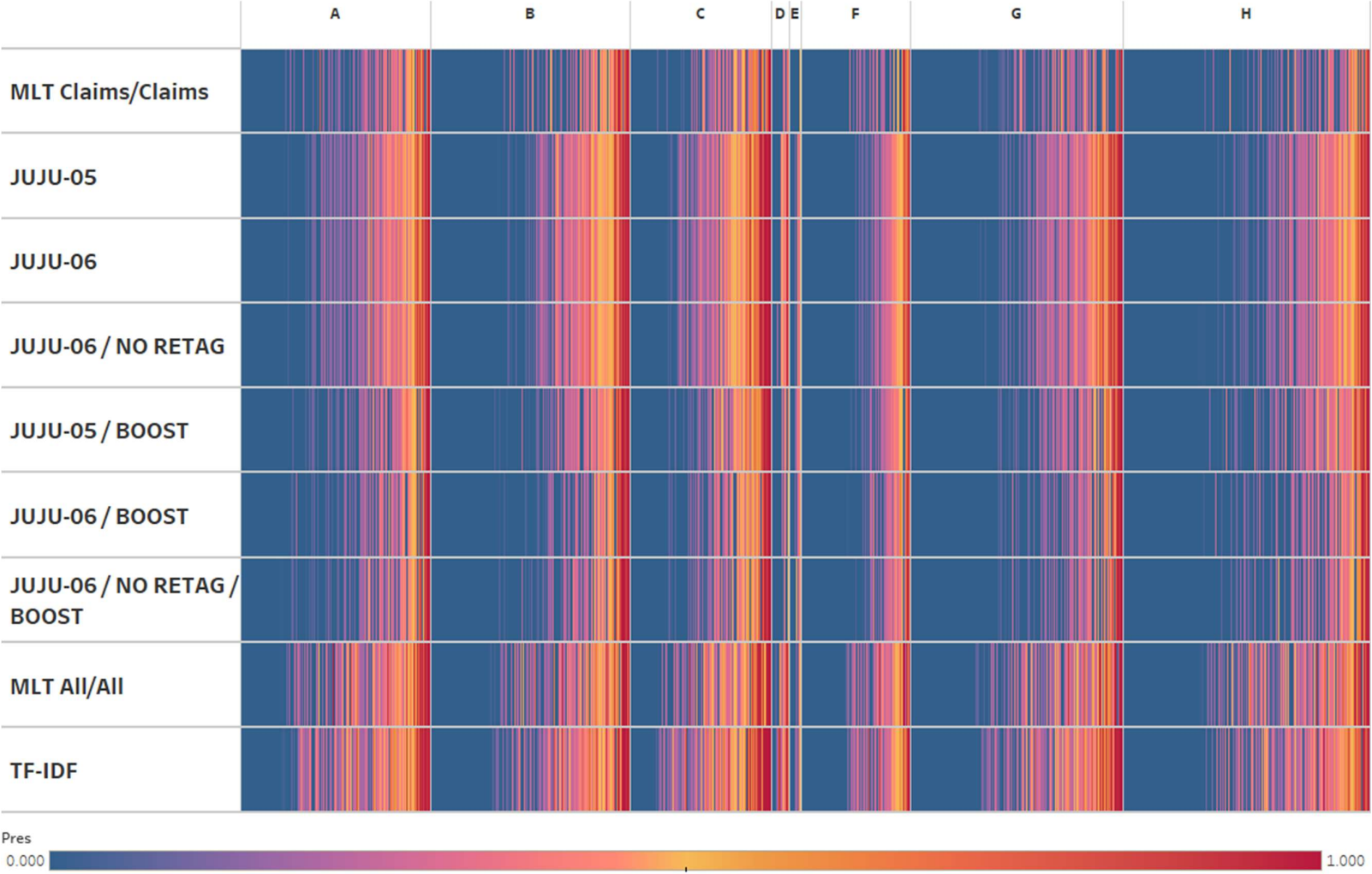

## 6.5 RANKING OF RELEVANT DOCUMENTS

We also study at what rank does the first relevant document occurs in the search results, with Figure 17 and Figure 18.

Figure 17 shows the difference between MLT and the best Juju algorithm for that matter. This study shows all cases when at least one relevant document was retrieved. The first line indicates for how many topics the first relevant document was in the corresponding rank category (ranks go from 1 to 1000, and is binned 10 by 10), the second line indicates a cumulative sum, this is the number of topics where the first hit appeared at any rank lower than the current one.

Figure 18 shows the same, only limited to cases where the first relevant document is appearing in the first 200 search results.

For example, Figure 17 reads like this:

With Juju-06, for 34% of the patent documents, the first relevant document in the search results appeared at a rank between 1 and 9. For 12% of the patent document, it was between 10 and 19. Cumulated, for 50% of patent documents, we have a first relevant result occurring in the first 20 search results (therefore, the median is 20). This is limited to the topics for which at least one relevant document was found.

For the comparison, the statistics are limited to the topics for which both algorithms generated a search result with a relevant document.

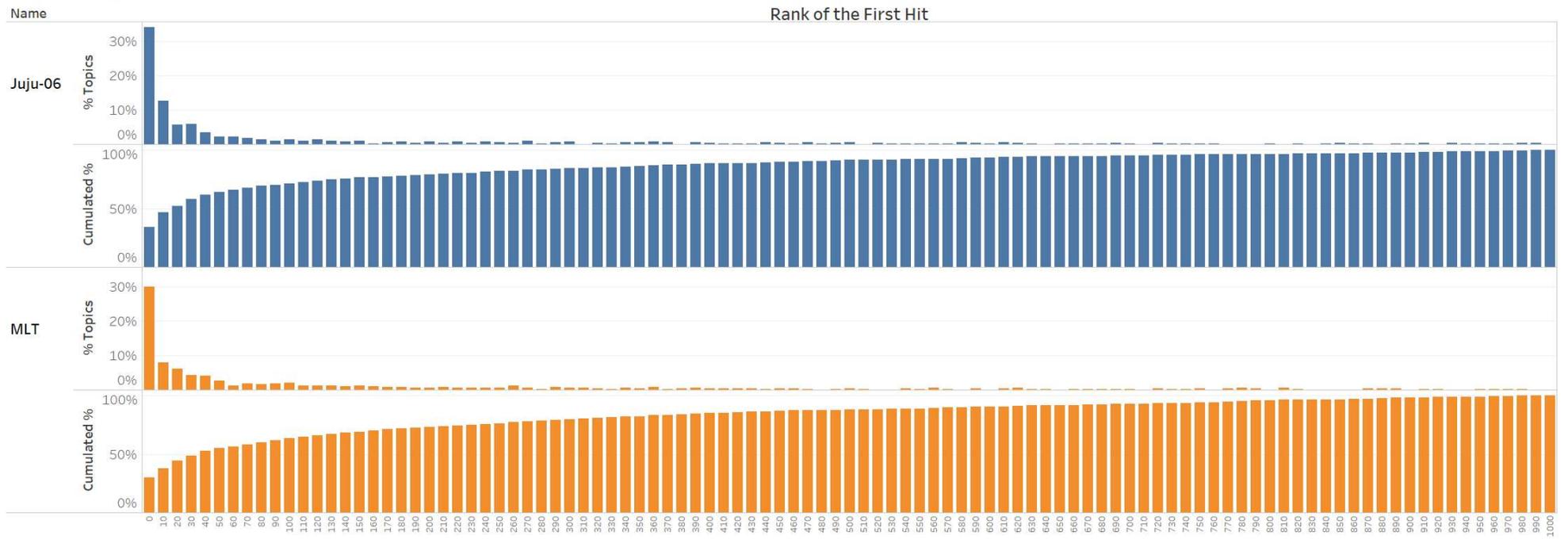

Figure 17 - Ranking of the First Hit

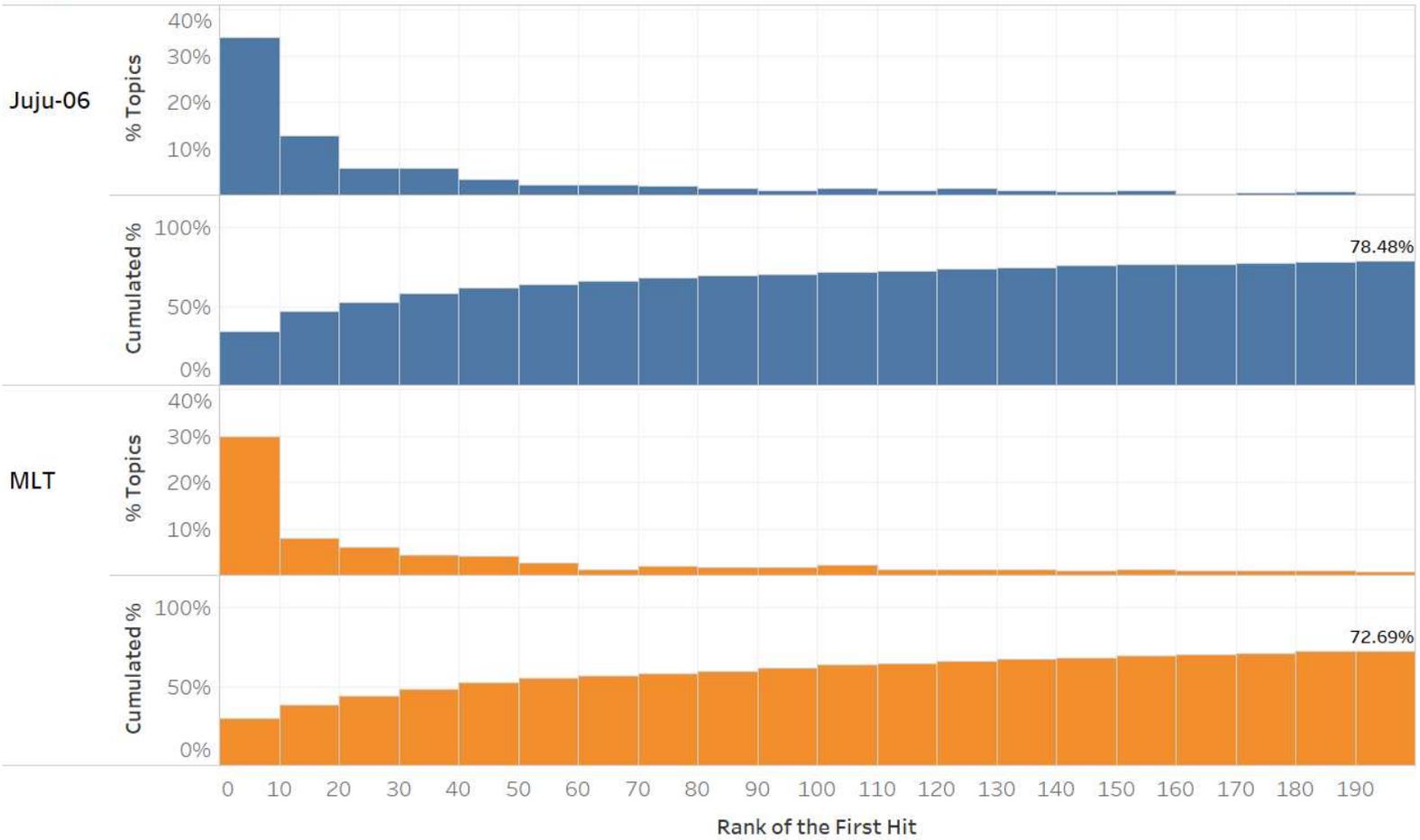

*Figure 18 - Ranking of the First Hit, Limited to 200*

We see that our algorithm brings more first hit in the first rank categories, therefore reducing the burden of search by presenting relevant results early.

This is illustrated by the higher percentage of topics with a rank for first hit in 0-10 or 10-20, but also we observe the median rank for the first hit is 40 for the MLT, while it is 20 for our algorithm. The 80% percentile is located at 220 for our algorithm, and 310 for the MLT.

This dimension is taken into account in PRES, and we can confirm that the higher PRES performance translates as well into a lower effort for search.

We also include visualizations similar to the heatmaps provided for PRES, only this time it represents the ranking of the first relevant document in the search results.

A better performance is represented by a "greener" drawing overall, indicating that more and more searches featured relevant document in the first 20 results.

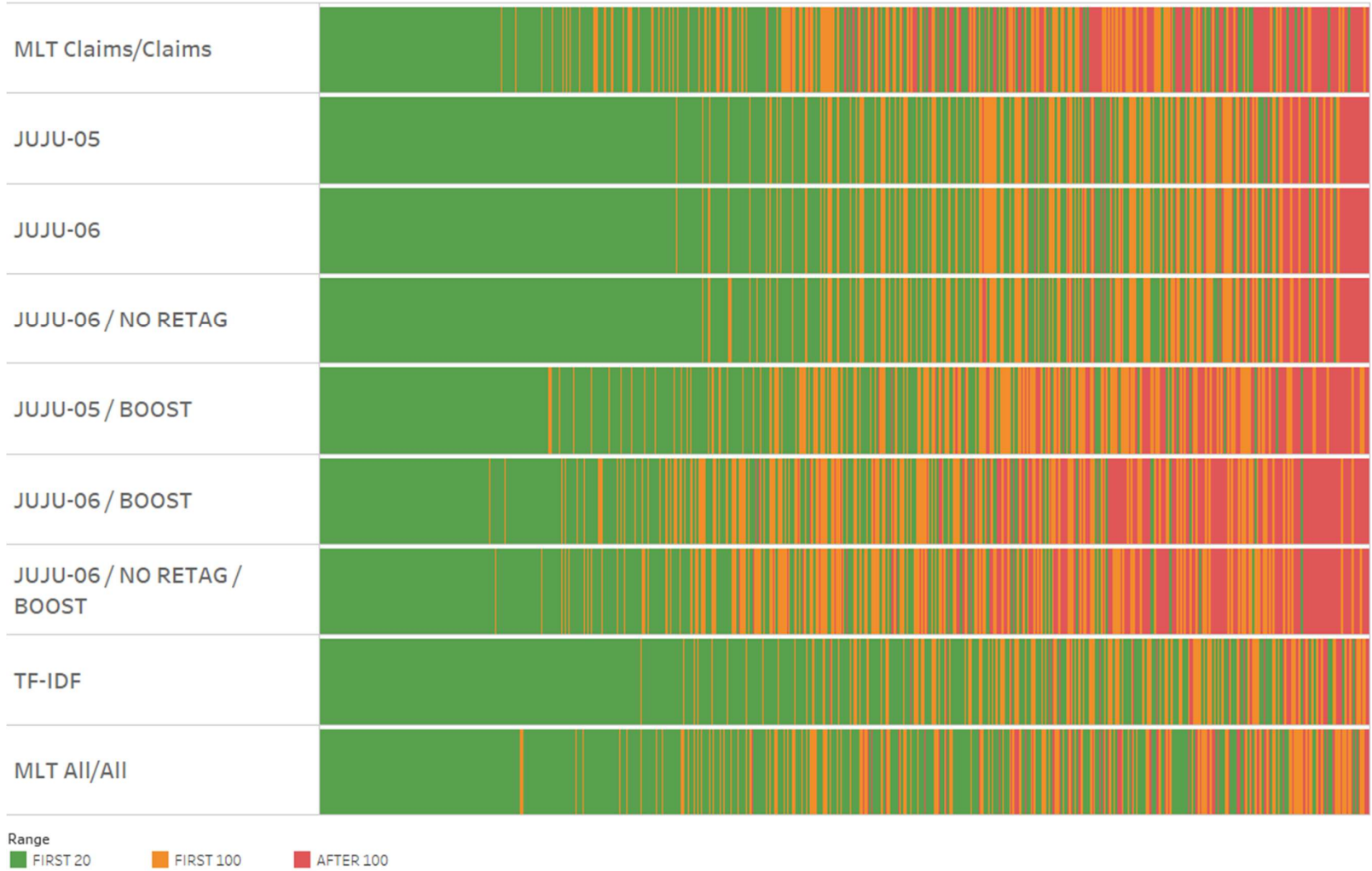

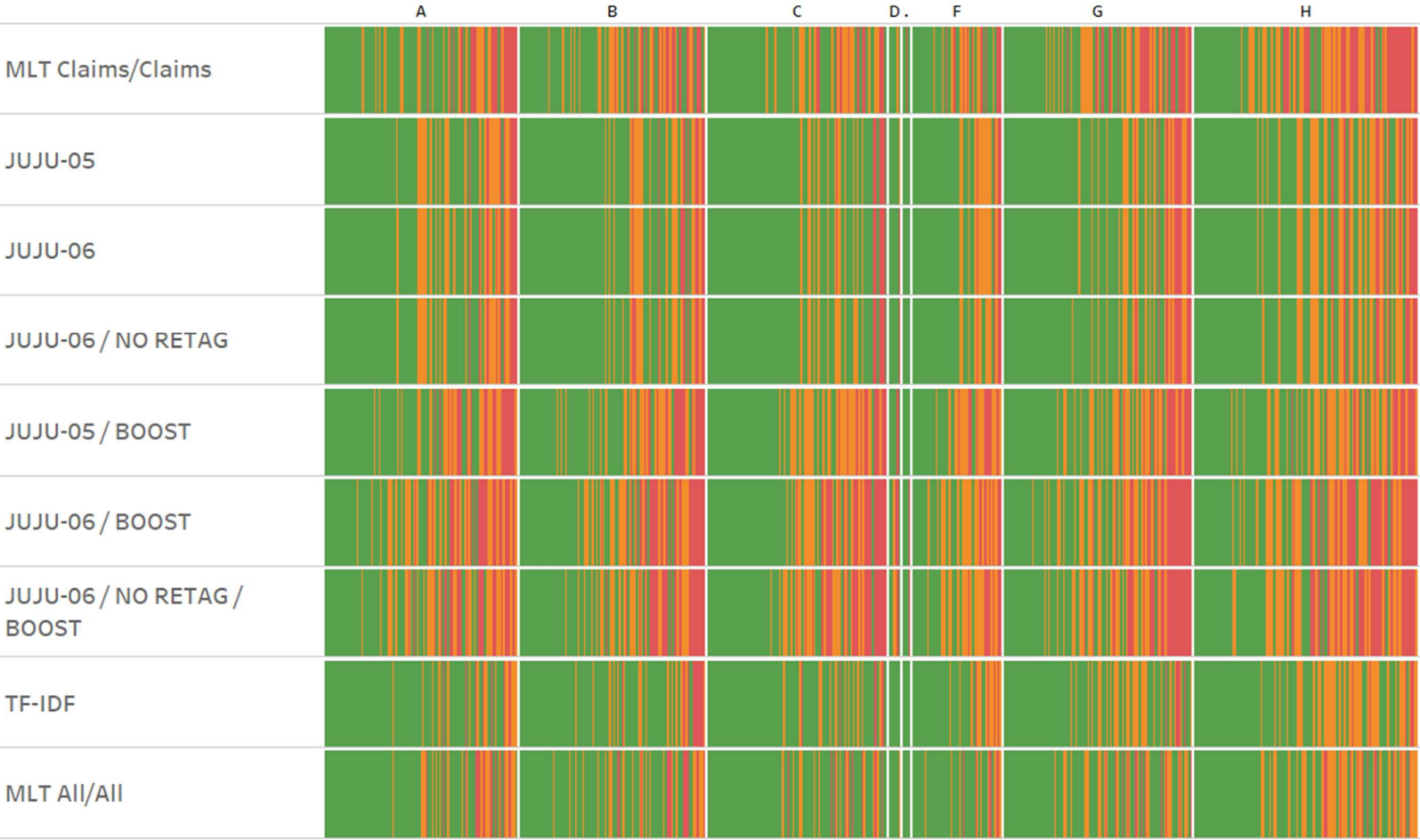

# 7 CONCLUSION

The newly developed algorithm outperforms the existing baseline of MLT significantly, in the context of searching "from claims to claims".

We also confirm this performance when focusing on PRES@50 or PRES@10, which indicates a capacity to get relevant results very early in the search results. We suggest to focus more on these metrics rather than Recall@1000 or PRES@1000, in order to deal with the reality of the usage of a search engine, where a thorough user will not continue to review results as far as after a rank estimated between 100 and 200 for most of the target population.

We have shown that the Specialization Trees can help to retrieve meaningful keywords from Patent Claims, as an exploitation of the way they are written out. In the context of the retrieval of Prior Art, we should expect to see only a limited improvement on the metrics, and we observe an improvement of 37% over the More Like This baseline on PRES@10, 45% on PRES@100, and 36% on PRES@1000, but in a setting that does not reflect the effective usage of the production systems.

The method is still likely to receive adjustments and improvements, a first one being the capacity to deal with Patent Documents in French and German languages. The EPO intends to continue the exploitation of these keywords and their location in the claims in order to better make use of the claims, in link with the description, in automated analyses. Furthermore, the EPO should integrate this keyword extraction method to existing algorithms in order to improve the performance of the production systems.

As observed in this research, the gap between "regular" literature and Patent Literature is not resolved in mainstream NLP tools. The training and use of a POS-tag corrector to allow for the incorrect classification of Verbs is one of many other challenges that would require a new generation of NLP tools to be trained specifically on patent claims and specifically on patent descriptions, in order to better fit the specificities of the languages.

The progresses made on keyword extraction also show the limitations of that method, as Recall and PRES seem to improve only asymptotically. We believe that further on, the possibilities of interactive searches and technologically assisted review should be tested in the setting of patent corpus navigation.